\documentclass[
aps,
prd,
superscriptaddress,
nofootinbib,
floatfix]
{revtex4}
\usepackage{graphicx,
longtable}
\usepackage{amsmath}
\DeclareMathOperator\erf{erf}
\begin{document}
\title{PINGU and the neutrino mass hierarchy: Statistical and systematic aspects}

%
\author{        F.~Capozzi}
\affiliation{   Dipartimento Interateneo di Fisica ``Michelangelo Merlin,'' 
               Via Amendola 173, 70126 Bari, Italy}%
\affiliation{   Istituto Nazionale di Fisica Nucleare, Sezione di Bari, 
               Via Orabona 4, 70126 Bari, Italy}
\author{        E.~Lisi}
\affiliation{   Istituto Nazionale di Fisica Nucleare, Sezione di Bari, 
               Via Orabona 4, 70126 Bari, Italy}
\author{        A.~Marrone}
\affiliation{   Dipartimento Interateneo di Fisica ``Michelangelo Merlin,'' 
               Via Amendola 173, 70126 Bari, Italy}%
\affiliation{   Istituto Nazionale di Fisica Nucleare, Sezione di Bari, 
               Via Orabona 4, 70126 Bari, Italy}
\begin{abstract}
The proposed PINGU project (Precision IceCube Next Generation Upgrade) is expected to collect 
$O(10^5)$ atmospheric muon and electron neutrino in a few years of exposure,
and to probe the neutrino mass hierarchy through its imprint on the event spectra 
in energy and direction. In the presence of nonnegligible and partly unknown shape systematics, 
the analysis of high-statistics spectral variations will face subtle challenges that 
are largely unprecedented in neutrino physics. We discuss these issues both on 
general grounds and in the currently envisaged PINGU configuration, where we find that possible 
shape uncertainties at the (few) percent level can noticeably affect the sensitivity to the 
hierarchy. We also discuss the interplay between the mixing angle $\theta_{23}$
and the PINGU sensitivity to the hierarchy.
Our results suggest that more refined estimates of spectral uncertainties are needed
in next-generation,  large-volume atmospheric neutrino experiments. 
\end{abstract}
\maketitle

\section{Introduction  \label{SecI}}

The discovery of atmospheric neutrino oscillations in the Super-Kamiokande (SK) experiment in 1998 \cite{SK98} marked the birth of what has now become the standard $3\nu$ mass-mixing scenario, where the neutrino flavor states $(\nu_e,\nu_\mu,\nu_\tau)$ are mixed with different states $(\nu_1,\nu_2,\nu_3)$ having definite masses $(m_1,m_2,m_3)$, via three mixing angles $(\theta_{12},\theta_{13},\theta_{23})$ and a possible CP-violating phase $\delta$ \cite{PDGR}. Several oscillation experiments have allowed to measure  
the three angles $\theta_{ij}$ (but not yet the phase $\delta$), as well as two independent squared mass differences among $\Delta m^2_{ij}= m^2_i-m^2_j$ \cite{PDGR}, which can be conventionally chosen as  $\delta m^2 = \Delta m^2_{21}>0$  and \cite{Sola,Glob}
\begin{equation}
\Delta m^2 = \pm \left| \frac{\Delta m^2_{31}+\Delta m^2_{32}}{2}\right| \ , 
\end{equation}   
where the sign of $\Delta m^2$ distinguishes the so-called normal ($+$) and inverted ($-$) neutrino mass hierarchy (NH and IH, respectively). Recent oscillation analyses with updated bounds on $\theta_{ij}$ and $\Delta m^2_{ij}$ are reported in \cite{Glob1,Glob2,Glob3}.  

Global data analyses have already shown a slight sensitivity to the hierarchy through various subsets of data, including reactor events \cite{Sola}, solar events \cite{Glob}, atmospheric events \cite{Scio} and, more recently, long-baseline accelerator events \cite{Glob1,Glob2,Glob3}. In particular, in the observable zenith distributions of atmospheric neutrinos, sub-horizon events carry information on sign$(\Delta m^2)$ via its interference with the effective neutrino potential $V$ in the Earth matter \cite{BlSm},  
\begin{equation}
V = \pm \sqrt{2}\,G_F\,N_e\ , 
\end{equation} 
where the upper (lower) sign refers to $\nu$ ($\overline\nu$),  $G_F$ is the Fermi constant, and $N_e$ is the electron number density. Even if $\nu$ and $\overline \nu$ are not separated, the atmospheric neutrino event rates are not symmetric under $\nu\leftrightarrow\overline\nu$ exchange \cite{Scio}, and thus hierarchy effects are not canceled. However, within the available data, such subleading effects have not emerged yet, being largely smeared out by the relatively coarse resolution in neutrino direction and energy \cite{Glob}. Indeed, in global three-neutrino $\chi^2$ fits to atmospheric data only, the difference between the two hierarchies amounted to a mere $\Delta \chi^2 \simeq 0.3$ in the pre-SK era \cite{Scio}, and it is still as small as $\Delta \chi^2 \simeq 0.9$ in the latest SK Collaboration analysis \cite{Wend}.

However, hierarchy effects may well emerge in next-generation, large-volume atmospheric neutrino experiments, provided that the event statistics and the resolutions in energy and angle are high enough, as recently emphasized in \cite{Smir}. In this context, the proposed ice-Cherenkov detector PINGU (Precision IceCube Next Generation Upgrade) \cite{PING}, which is expected to collect O($10^5$) neutrino events in a few years, appears to be a very promising project, and is currently under extensive investigation. Comparable goals might be reached in the proposed water-Cherenkov detector ORCA (Oscillation Research with Cosmics in the Abyss) \cite{ORCA} and Hyper-Kamiokande \cite{HYPE}, while the ICAL-INO project (Iron CALorimeter at the India-based Neutrino Observatory) \cite{ICAL} aims at increasing the sensitivity to the hierarchy with a different approach ($\nu$ and $\overline\nu$ event separation). We refer the reader to \cite{Cahn} for a recent survey of possible approaches to the hierarchy discrimination, by using atmospheric (and other) neutrino sources.   

In this paper we focus on PINGU, taken as a case study for analyzing the atmospheric neutrino sensitivity to the hierarchy with very high statistics. Several works have already addressed this task \cite{Fran,Wint,Blen,Colo,Hagi}, showing that PINGU can reach a sensitivity of at least a few standard deviations in a few years, especially in favorable conditions for large matter effects (i.e., for normal hierarchy and $\theta_{23}$ in the second octant). While we confirm this broad picture with an independent analysis, we try to bring to surface several subtle issues related to the calculation of spectral shapes and to the estimate of their systematic uncertainties, which represent very peculiar challenges of high-statistics data sets. Although such issues have already emerged in other fields, such as parton distribution fitting \cite{Pump,Ball} and precision cosmology \cite{Kitc,Tayl}, they have only been touched upon in neutrino physics (see, e.g., \cite{Kaji}) and deserve renewed attention. In addition, we elucidate the interplay between the determination of the hierarchy and the measurement of $\theta_{23}$ in PINGU.  

Our work is structured as follows. In Sec.~II we describe our calculation of spectral event rates in PINGU and their binning in energy and (zenith) angle. In Sec.~III we describe some general features of the statistical analysis of PINGU prospective data, including oscillation parameter uncertainties and normalization systematics, as well as other possible (correlated and uncorrelated) shape systematics, which play a relevant role in the limit of high statistics. In Sec.~IV we analyze in detail 
the impact of shape systematics on the PINGU sensitivity to the hierarchy. Finally, in Sec.~V we discuss the interplay between 
the hierarchy and $\theta_{23}$ constraints. We conclude with a brief summary and  
perspectives for further work in Sec.~VI.

\section{Calculation of energy-angle spectra in PINGU  \label{SecII}}

Our calculation of energy-angle spectra is described below. The reader not interested in  details
may jump to the last two subsections, where we discuss features of the spectra which are relevant for the
subsequent statistical analysis.

\subsection{Notation}

The following notation is used hereafter:
$$
\begin{array}{lcl}
\alpha						&=&\text{flavor index $(\mu,e)$ of $\nu$ or $\overline\nu$} ,\\
N^\alpha					&=&\text{number of}\ \nu_\alpha +\overline\nu_\alpha\ \text{events} ,\\
E',\,\theta'				&=&\text{true neutrino energy and zenith angle} ,\\
E,\,\theta 					&=&\text{reconstructed neutrino energy and zenith angle} ,\\
r^\alpha_E(E,E')					&=&\text{energy resolution function} ,\\
r^\alpha_{\theta}(\theta,\theta')	&=&\text{angular resolution function} ,\\
T							&=&\text{detector livetime} , \\
\rho V^\alpha_{\text{eff}}(E')     &=& \text{effective detector mass at energy}\ E' ,\\ 
d^2\Phi^{\alpha}/(d\cos\theta'dE')&=&\text{double differential neutrino flux}\ (\overline\Phi\ \text{for}\ \overline\nu) ,\\
\Phi^{\alpha}/\Phi^{\beta}  &=& \text{ratio of double differential neutrino fluxes} ,\\
\sigma^\alpha_{\text{CC}}(E') 		&=& \text{neutrino charged-current cross section}\ (\overline\sigma\ \text{for}\ \overline\nu) ,\\
P_{\alpha\beta}(\theta',E')	&=& \text{oscillation probability of}\ \nu_{\alpha}\rightarrow\nu_{\beta}\ (\overline P\ \text{for}\ \overline\nu) ,\\
\left[\theta_i, \theta_{i+1}\right] 	&=& \text{range of\ }i\text{-th angular bin} ,\\
\left[E_j, E_{j+1}\right] 	&=& \text{range of\ }j\text{-th energy bin} .
\end{array}
$$
Note that $\theta'/\pi=1$ and $0.5$ correspond to vertically upgoing and horizontal neutrino directions, respectively.

\subsection{Neutrino production and propagation}

Concerning the unoscillated atmospheric neutrino fluxes, we assume the  azimuth-averaged values of $d^2\Phi^\alpha/(d\cos\theta' dE')$ calculated at South Pole in \cite{HKKM}. These fluxes are modulated by the oscillation probabilities $P_{\alpha\beta}$ during neutrino propagation from the source (assumed to be a layer at $h=15$~km in the atmosphere) to the detector. For sub-horizon trajectories, matter effects are calculated up to a second-order Magnus expansion in each Earth shell as in \cite{Magn}, along an accurate model of the electron density profile \cite{Eart}. 

The probabilities depend, in general, on all the oscillation parameters ($\delta m^2,\,\pm \Delta m^2,\,\theta_{12},\,\theta_{13},\,\theta_{23},\,\delta$).
Whenever we need to fix ``true'' oscillation parameters to calculate an input spectrum for subsequent fits, we assume the following representative input values \cite{Glob1}: 
\begin{eqnarray}
\label{Dm2} |\Delta m^2|_{\text{true}} &=& 2.40\times 10^{-3}\ \text{eV}^2\ , \\
\label{dm2} \delta m^2|_{\text{true}} &=& 7.54\times 10^{-5}\ \text{eV}^2\ , \\
\label{q13} \sin^2\theta_{13}|_{\text{true}} &=& 0.0237 \ ,\\
\label{q12} \sin^2\theta_{12}|_{\text{true}} &=& 0.308 \ ,\\
\label{dcp} \delta|_{\text{true}} &=& 3\pi/2 \ .
\end{eqnarray}
The parameter $\theta_{23}$ is treated differently since, as discussed later, it induces large variations in the PINGU sensitivity to the hierarchy. Unless stated otherwise, we assume by default that it can take any true value in the range 
\begin{equation}
\label{q23} \sin^2\theta_{23}|_{\text{true}} \in [0.4,\,0.6]\ ,
\end{equation}
which covers both octants of $\theta_{23}$ (as currently allowed at $\sim2\sigma$ level in \cite{Glob1}). Reconstructed values of $\sin^2\theta_{23}$ are allowed to extend beyond this range in fits. The effects of prior ranges different from Eq.~(\ref{q23}) are discussed in Sec.~V.

\vspace*{-3mm}
\subsection{Neutrino detection}

Concerning the PINGU detector, we basically assume the preliminary characterization reported in \cite{PING}. We approximate the effective detector mass
$\rho V^{\alpha}_{\text{eff}}(E')$ (i.e., the ice density times the effective volume for $\alpha=\mu,e$) by interpolating the $E\geq 1$~GeV histograms in Fig.~6 of \cite{PING} with the following (smooth and monotonic) empirical  functions, 
\begin{eqnarray}
\rho V^{\mu}_{\text{eff}}(E') &=& 3.33 \left(1-e^{-0.287(E'-E'_{\text{thr}})}\right)\ ,\\
\rho V^{e}_{\text{eff}}(E') &=& 3.44 \left(1-e^{-0.294(E'-E'_{\text{thr}})}\right)\ ,
\end{eqnarray}
where $[\rho V^{\alpha}_{\text{eff}}]=\text{MTon}$, $[E']=\text{GeV}$, and the effective threshold has been set at 
\begin{equation}
E'_{\text{thr}}=1~\text{GeV}\ .
\end{equation}
The ratio of $\rho V^{\alpha}_{\text{eff}}$ to the proton mass $m_p$ provides the effective number of target nucleons.

Total charged current (CC) cross sections $\sigma^\mu_\text{CC}$ and 
$\bar{\sigma}^\mu_\text{CC}$ for $E'\geq E'_{\text{thr}}$ are extracted from Fig.~14 of \cite{PING}. They are the sum of three contributions: $(i)$ deep inelastic, $(ii)$ quasi-elastic and $(iii)$ resonant scattering, where the first
one dominates for $E'$ above a few GeV. For simplicity, we assume that the total CC cross section for $\nu_e$ is identical to the one for $\nu_{\mu}$ 
at any $E'\geq E'_{\text{thr}}$ (and similarly for antineutrinos),
\begin{equation}
\sigma^e_\text{CC}(E')=\sigma^\mu_\text{CC}(E')\equiv\sigma_\text{CC}(E')\ .
\end{equation}

The resolution functions are extracted from the 2-dimensional histograms in Fig.~14 of \cite{PING} in digitized form \cite{Resc}. In particular, in each $x$-axis bin having median true energy $E'$ therein, we fit the histogram contents with gaussian functions having widths $\sigma_E(E')$ and $\sigma_\theta(E')$,
\begin{eqnarray}
r^\alpha_E(E,E') &=& \frac{1}{\sqrt{2\pi}\sigma^\alpha_E(E')}\exp \left[-\frac{1}{2}\left(\frac{E-E'}{\sigma^\alpha_E(E')}\right)^2\right]\ ,
\label{resole}
\\
r^\alpha_\theta(\theta,\theta') &=& \frac{1}{\sqrt{2\pi}\sigma^\alpha_\theta(E')}\exp \left[-\frac{1}{2}\left(\frac{\theta-\theta'}{\sigma^\alpha_\theta(E')}\right)^2\right]\ .
\label{resolq}
\end{eqnarray}
The resulting collection of widths $\sigma^\alpha_E(E')$ and $\sigma^\alpha_\theta(E')$  in each bin of $E'$ are finally fitted with the following (smooth and monotonic) empirical functions:
\begin{eqnarray}
\sigma^\mu_E/E' &=& 0.266/(E'^{0.171}-0.604)\ ,\\
\sigma^e_E/E' &=& 0.369/(E'^{0.247}-0.508)\ ,\\
\sigma^\mu_\theta &=& 3.65/(E'^{1.05}+5.00)\ ,\\
\sigma^e_\theta &=& 1.88/(E'^{0.823}+1.93)\ ,
\end{eqnarray}
where $[E']=\text{GeV}$ and $[\sigma_\theta]=\text{rad}$. These approximations 
capture the main features of PINGU as described in \cite{PING}. 

\begin{figure}[t]
\centering
\includegraphics[height=4.9cm]{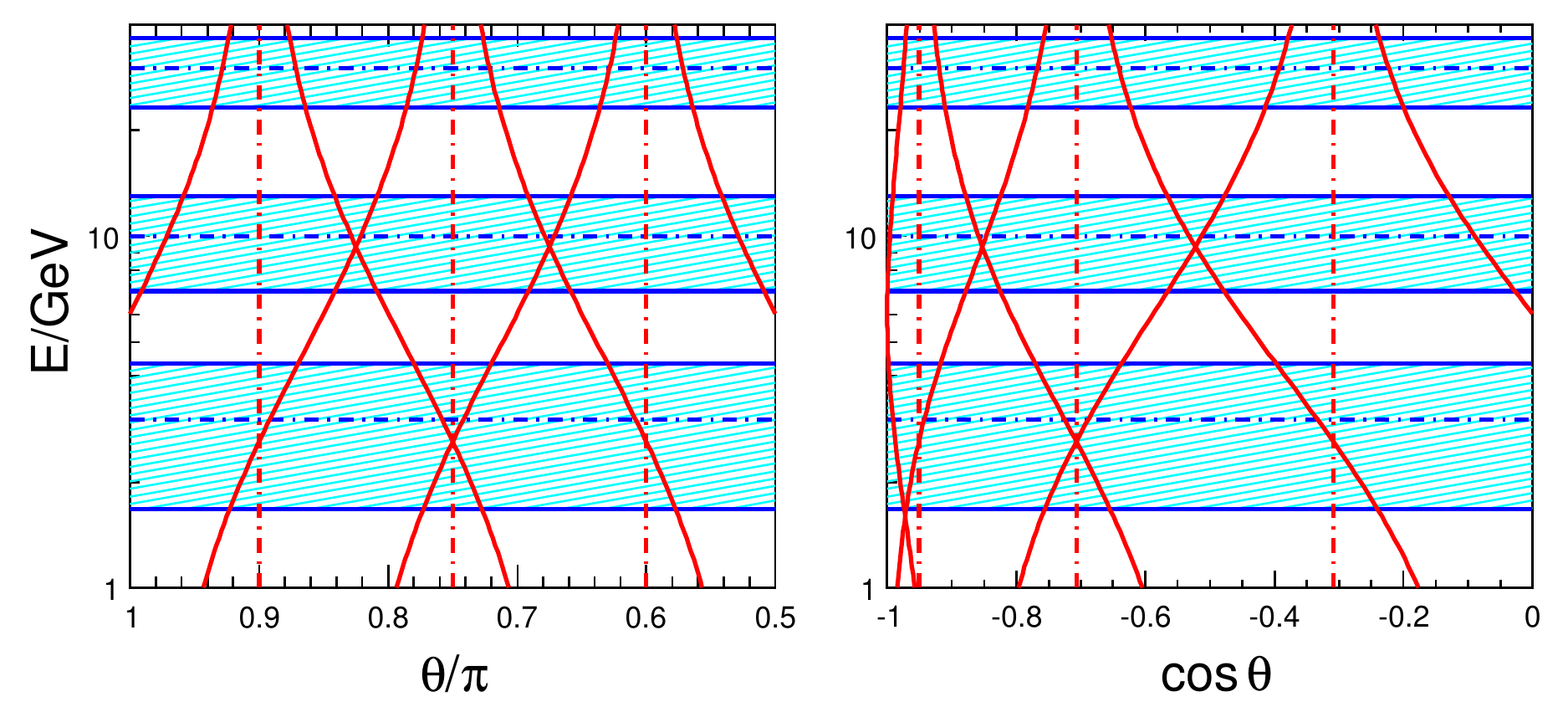}
\vspace*{-3mm}
\caption{Widths (at $\pm1\sigma$) of the resolution functions in energy and angle for $\nu_\mu$-like events in PINGU, in terms of logarithmic energy versus the zenith angle (left plot) or its cosine (right plot). See the text for details.
\label{Fig_01}}
\end{figure}

Figure~1 shows the $\pm 1\sigma$ resolution bands for $\nu_\mu$ 
in the plane charted by $\log_{10}(E'/\text{GeV})$ versus $\theta'/\pi$ (left plot) or versus
$\cos\theta'$ (right), in the intervals $E'\in[1,\,40]$~GeV and $\theta\in[0.5,\,1]$.%
\footnote{As usual, upgoing events ($\theta/\pi\sim 1$) correspond to the left of the zenith scale, and horizontal events ($\theta/\pi\sim 0.5$) to the right. }
The horizontal bands correspond to 
$\pm \sigma^\mu_E(E')$ 
for $E'=3,$~10, and 30~GeV, while the three curved, vertical bands correspond to $\pm \sigma^\mu_\theta(E')$ for $\theta'/\pi=0.6,$~0.75 and~0.9. The resolution functions  largely smear any spectral feature when passing from true to reconstructed variables, $(E',\,\theta')\to(E,\,\theta)$, the more the lower the energy. 

Figure~1 also illustrates three advantages of using the zenith angle rather than its cosine. The first is that the angular resolution bands (which provide a rough idea of the appropriate zenithal binning) are obviously symmetric in $\theta'$ (left plot) but not in $\cos\theta'$ (right plot), where they are squeezed towards the upgoing directions ($\cos\theta'\to -1$). The second is that, compared with the full sub-horizon range
$\theta'/\pi\in [0.5,1]$, the interesting angular fraction subtending the dense Earth core ($\theta'/\pi\in[0.816,1]$) is as large as 36.8\%, while it would be squeezed by a factor of about two (16.2\%) in terms of $\cos\theta'$. The third is that, by using $\cos\theta'$, one would expand the nearly horizontal part of the zenith spectrum ($\cos\theta'\gtrsim -0.5$) which, despite being weighted by higher atmospheric fluxes \cite{HKKM}, is less interesting for hierarchy discrimination, due to smaller matter effects at shallow depth in the Earth's mantle. 

Concerning the energy, we remind that in normal (inverted) hierarchy, matter effects for neutrinos (antineutrinos) are particularly enhanced around $E'\sim 2.5$--3~GeV and  $E'\sim 6$--10~GeV (corresponding to resonance effects in the core and in the mantle, respectively), as well as for intermediate energies where mantle-core interference effects occur (see \cite{PDGR,Smir} and refs.\ therein). Although the low-energy range $E'\sim O(1-10)$~GeV contains most of the hierarchy
``signal,'' it is useful to extend the analysis to few tens of GeV (or more), for at least two reasons: (1) the high-energy spectrum is better experimentally resolved and is largely hierarchy-independent, so it can help to ``fix'' some floating parameters in the fits; (2) due to the relatively poor energy resolution
at low energy, hierarchy effects may ``migrate'' well above $\sim 10$ GeV in reconstructed energy. In any case, to avoid ``squeezing'' the most relevant low-energy range, it is useful to adopt a logarithmic energy scale, as in Fig.~1.  Summarizing,  we shall use the zenith angle (instead of its cosine) and a logarithmic energy scale
for representing PINGU event spectra.

\vspace*{-3mm}
\subsection{Unbinned event spectra}

In atmospheric neutrino experiments, the detected neutrino events are usually organized in terms of energy and zenith angle (or related variables). 
The double differential spectra of $N^\alpha$ events induced in PINGU by both $\nu_\alpha$ and $\overline\nu_\alpha$, as a function of the true energy $E'$ and zenith angle $\theta'$, can be cast in the form
\begin{equation}
\frac{d^2N^\alpha}{d\cos\theta'dE'} = \left[
2\pi \, T\,\frac{\rho V^\alpha_{\text{eff}}(E')}{m_p}\sigma_{\text{CC}}(E')\frac{d^2\Phi^{\alpha}(\theta',E')}{d\cos\theta'dE'}
\right] P^\alpha(\theta',E')\ ,
\label{2Dtrue}
\end{equation}
where the prefactor in square brackets does not depend on the oscillation parameters, while the last factor $P^\alpha$ is a linear combination of the relevant oscillation probabilities:
\begin{equation}
P^\alpha = \left[ P_{\alpha\alpha} + \frac{\Phi^\beta}{\Phi^\alpha}P_{\beta\alpha} \right] + 
\left[ \frac{\overline\Phi^\alpha}{\Phi^\alpha}\frac{\overline\sigma_{\text{CC}}}{\sigma_{\text{CC}}}\overline P_{\alpha\alpha} + \frac{\overline \Phi^\beta}{\Phi^\alpha}\frac{\overline\sigma_{\text{CC}}}{\sigma_{\text{CC}}} \overline P_{\beta\alpha} \right]  \ ,
\label{Palpha}
\end{equation}
where $\beta\neq \alpha$ and the first (second) term in brackets is due to $\nu$ ($\overline\nu$), respectively. 
In Eq.~(\ref{2Dtrue}) we have assumed a priori azimuthal averaging, hence the $2\pi$ factor; see \cite{Hagi} for an approach including azimuth dependence. This issue and other related approximations will be discussed in Sec.~II~G.

The spectra in terms of reconstructed variables $(E,\,\theta)$ are obtained by convolving the ones in Eq.~(\ref{2Dtrue}) with the  resolution functions,
\begin{equation}
\frac{d^2N^\alpha}{d\theta dE} = \int_{0}^{2\pi}\sin\theta' d\theta' r_{\theta}^\alpha(\theta,\theta')\int_{E'_{\text{thr}}}^\infty dE' r^\alpha_E(E,E') \frac{d^2N^\alpha}{d\cos\theta'dE'}\ , 
\label{2Dreco}
\end{equation}
where the change of variable $\cos\theta'\to\theta'$ has been applied, as discussed at the end of the previous subsection.

\subsection{Binned event spectra}

The number of events $N^\alpha _{ij}$ in the $ij$-th bin is obtained by integrating the r.h.s.\ of Eq.~(\ref{2Dreco}) over the bin area 
$[\theta_i,\,\theta_{i+1}]\otimes [E_j,\,E_{j+1}]$. 
By changing the integration order (see \cite{Faid,Capo}), the resulting quadruple integration can be reduced to a double one:
\begin{eqnarray}
N_{ij}^\alpha &=& \int_{\theta_i}^{\theta_{i+1}}d\theta \int_{E_j}^{E_{j+1}}dE \int_{0}^{2\pi}\sin\theta' d\theta' r_{\theta}^\alpha(\theta,\theta')\int_{E'_{\text{thr}}}^\infty dE' r_E(E,E') \frac{d^2N^\alpha}{d\cos\theta'dE'}  \\
&=& \int_{0}^{2\pi}\sin\theta' d\theta' \int_{E'_{\text{thr}}}^\infty dE' w^\alpha_i(\theta')\,w^\alpha_j(E') \frac{d^2N^\alpha}{d\cos\theta'dE'}\ ,
\label{double}
\end{eqnarray}
where the  functions $w^\alpha_n(x)$ are defined, for $(n,x')=(i,\theta')$ and $(j,E')$, as:  
\begin{equation}
w_n(x') = \frac{1}{2}\erf\left(\frac{x_{n+1}-x'}{\sqrt{2}\sigma_x^\alpha}\right)-\frac{1}{2}\erf\left(\frac{x_{n}-x'}{\sqrt{2}\sigma_x^\alpha}\right)\ ,
\label{window}
\end{equation}
with $\erf(x)$ defined as in \cite{Inte}, 
\begin{equation}
\erf(x)=\frac{2}{\sqrt{\pi}}\int_0^x dte^{-t^2}\ .
\label{erf}
\end{equation}
In the limit of perfect resolution ($\sigma_x^\alpha\to 0$), the curve $w_n(x')$ becomes a top-hat function in the interval $[x_n,\,x_{n+1}]$. For finite resolution, the ``top-hat'' shape is  smeared and extends beyond this interval for a few $\sigma_x^\alpha$'s. However, for numerical purposes, $w_n(x')$ practically vanishes beyond $[x_n- 4\sigma_x^\alpha,\,x_{n+1}+4\sigma_x^\alpha]$,
so that the double integral domain in Eq.~(\ref{double}) can be just taken as the $ij$-th bin range ``augmented'' by 
$\pm4\sigma^\alpha_\theta$ and $\pm4\sigma^\alpha_{E}$. 

As previously argued, we actually adopt a logarithmic energy variable, 
\begin{equation}
\lambda = \log_{10}(E'/\text{GeV})\ ,
\label{lambda}
\end{equation}
(and similarly for the reconstructed energy $E$), so that 
\begin{equation}
N_{ij}^\alpha =  \int_{0}^{2\pi}\sin\theta' d\theta' \int_{0}^\infty d \lambda\, E' \ln(10)\,  w^\alpha_i(\theta')\,w^\alpha_j(E') \frac{d^2N^\alpha}{d\cos\theta'dE'}\ .
\label{doublelog}
\end{equation}
Finally, we consider reconstructed energies in the interval $E\in [1,\,40]$~GeV, namely, in the logarithmic range $\log_{10}(E/\text{GeV})\in [0,\,1.6]$, that we  divide into 16 bins. We also divide the range of sub-horizon reconstructed angle, $\theta/\pi=[0.5,\,1]$, into 10 bins.  With this choice, the bin widths are smaller than the typical resolution widths in Fig.~1 (so as to avoid additional smearing from binning), but large enough to contain a significant number of events after a few years of exposure (so as to apply Gaussian, rather than Poissonian, statistics). The calculation of $N_{ij}^\alpha$ via Eq.~(\ref{doublelog}) is performed through Gauss quadrature routines, which have been checked to yield numerically stable results up to the third significant figure, even in bins where the integrand  oscillates rapidly via $P^\alpha$.

\subsection{Qualitative discussion of spectral shapes}

Figure~2 shows the main ingredients of typical PINGU spectra calculations (first three couples of panels from the left) and final spectra (last couple of panels on the right), where the upper and lower panels refer to muon events ($\alpha=\mu$) and to electron events ($\alpha=e$), respectively. The adopted ranges and bins have been discussed in the previous subsection. 
For definiteness, we have assumed normal hierarchy (NH), $\sin^2\theta_{23}=0.5$, and the remaining oscillation parameters as in Eqs.~(\ref{Dm2})--(\ref{dcp}); in any case, the graphical results would appear qualitatively similar for different choices. The units and the color scale are arbitrary: in each panel, the 
darkest color corresponds to the bin with maximum contents, while lighter shades refer to lower contents, down to total white for almost empty bins.%
\footnote{Concerning the absolute event rates, for the specific oscillation parameters chosen in Fig.~2, we estimate a total of $1.9\times 10^4$ muon and $1.4\times 10^4$ electron events per year in PINGU.  The total statistics can thus reach $O(10^5)$ events in a few years, as already noted.}

The leftmost panels in Fig.~2 show the product $V_{\text{eff}}^\alpha \Phi^\alpha\sigma^\alpha_{\text{CC}}$, namely, the oscillation-independent prefactor in Eq.~(\ref{2Dtrue}). This prefactor is suppressed at high energy by $ \Phi^\alpha\sigma^\alpha_{\text{CC}}\sim E^{-2}$, and at low energy by the small value of 
$V^\alpha_{\text{eff}}$, with a maximum in the few GeV range, which is interesting for matter effects. Unfortunately, the atmospheric neutrino flux $\Phi^\alpha$ peaks at the right energy but in the wrong direction, i.e., at the horizon ($\theta/\pi\to 0.5$), where matter effects vanish. In this sense, atmospheric neutrinos are not ``optimal'' for seeking hierarchy effects, which occur mainly in a tail (rather than at the peak) of the event spectrum.

The next couple of panels in Fig.~2 show the oscillation-dependent factor $P^\alpha$ in Eq.(\ref{Palpha}). In the upper panel, the factor $P^\mu$ shows large variations, whose shape is reminiscent of the oscillograms related to $\nu_\mu$ disappearance \cite{Malt}. In particular, a large disappearance ``valley'' (the first oscillation minimum) extends from the upper left corner to the lower right margin of the panel. Conversely, in the lower panel, the factor $P^e$  shows much milder variations, since the $\nu_e$ disappearance and appearance probabilities are largely suppressed by the smallness of $\sin^2\theta_{13}$. 

The third couple of panels shows the binned product of the factors $V_{\text{eff}}^\alpha \Phi^\alpha\sigma^\alpha_{\text{CC}}$ and $P^\alpha$, which is proportional to the unsmeared spectrum of events in Eq.~(\ref{2Dtrue}). An oscillatory structure is still visible in the central part of each panel, 
i.e., for slanted trajectories and for $E\sim\text{few}$~GeV. These structures, however, are largely suppressed around the vertical upgoing direction, 
where the atmospheric flux is lower. 

Finally, the rightmost couple of panels shows the observable, smeared spectra of $\mu$ and $e$ events, including  resolution effects as in Eq.~(\ref{doublelog}). The oscillating structures appear to be largely suppressed, except for remnants of the large $\nu_\mu$ disappearance valley, carrying the dominant information about the oscillation parameters $(|\Delta m^2|,\,\sin^2\theta_{23})$. 
The smeared spectra  in inverted hierarchy (not shown) would be visually indistinguishable from the ones in normal hierarchy in Fig.~2. 
Indeed, as discussed in the next Section, hierarchy effects emerge as relatively smooth and subdominant modulations, at the level of a few percent,
in the left tail of the zenith spectra and for intermediate energies. It is thus imperative to assess how accurately we know the {\em shapes\/} of the  neutrino event spectra in PINGU and, in general, in large-volume atmospheric experiments. 
 
\begin{figure}[t]
\hspace*{-3mm}
\includegraphics[height=10.9cm]{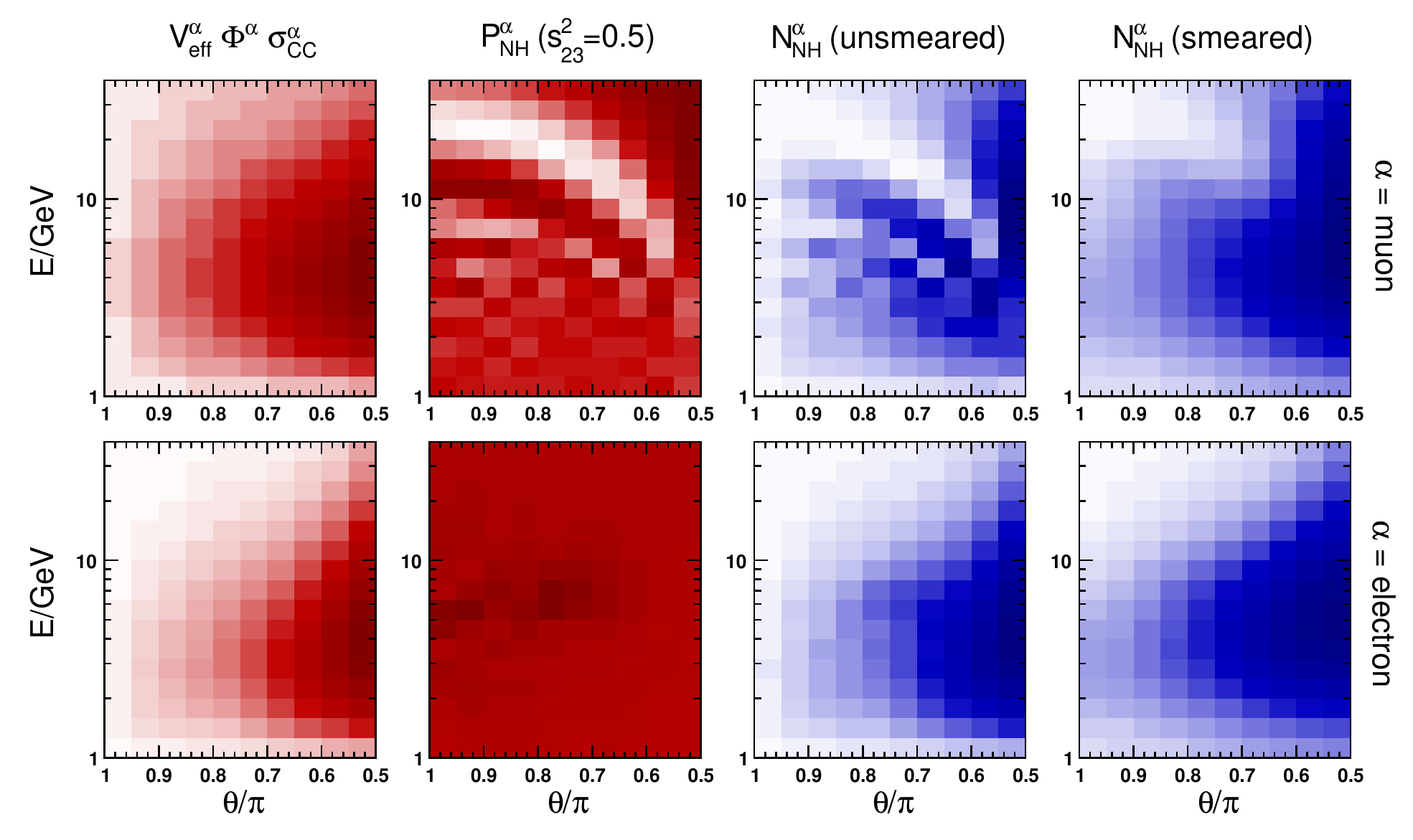}
\caption{Breakdown of the main ingredients of typical muon (upper panels) and electron (lower panels) event spectra in energy and angle in PINGU. From left to right: oscillation-independent factor $V_{\text{eff}}^\alpha \Phi^\alpha\sigma^\alpha_{\text{CC}}$, oscillation-dependent factor  $P^\alpha$, their product (unsmeared event spectrum), and final results (smeared event spectrum). See the text for details.
\label{Fig_02}}
\end{figure}

\subsection{Qualitative discussion of shape uncertainties}
\vspace*{-2mm}

In a few years, PINGU is expected to collect $O(10^5)$ events which, distributed in $O(10^2)$ bin contents $N^\alpha_{ij}$, can constrain the energy-angle spectral {\em shape\/} with a statistical accuracy $1/\sqrt{N^\alpha_{ij}}\sim\text{few}\%$. However, the shape may also be affected by comparable
systematic errors $\delta N^\alpha_{ij}/N^\alpha_{ij}$, stemming from 
the oscillation-independent factors ($V_{\text{eff}}^\alpha \Phi^\alpha\sigma^\alpha_{\text{CC}}$), the oscillation-dependent ones ($P^\alpha$), the resolution functions ($r^\alpha_{E,\theta}$), and the event integration into bins.%
\footnote{Although some spectral shape errors have been previously considered in the literature \protect\cite{Smir,PING,Fran,Wint,Blen,Colo,Hagi}, 
we think it useful to present a more extended and self-contained discussion herein.}

The effective detector volume $V^\alpha_{\text{eff}}$, which carries information about the event detection and reconstruction efficiency, has been estimated in PINGU via Monte Carlo (MC) simulations \cite{PING}. In particular, Fig.~6 therein shows its $E'$ dependence, with 
fluctuations suggestive of finite MC statistics. In general, the event collection efficiency depends also on $\theta'$ (due to the cylindrical configuration) \cite{Mena} and, to some extent, on the azimuth angle $\phi'$ (due to unavoidable anisotropies of the detector and inhomogeneities of the ice). Apart from  normalization and finite MC statistics errors, it is reasonable to think that the $V^\alpha_{\text{eff}}$'s are 
affected by {\em shape\/} uncertainties  $\delta V^\alpha_{\text{eff}}(E',\,\theta',\,\phi')$ at the (few) percent 
level, possibly different for $\mu$ and $e$,  especially in the relevant range $E\sim\text{few}$~GeV, where the $V^\alpha_{\text{eff}}$'s  change rapidly.%
\footnote{For the sake of comparison, in 
a laboratory neutrino experiments such as Borexino \protect\cite{Bore}, deviations of the effective 
fiducial volume from its quasi-spherical shape are estimated to be at the few percent level from point to point [$O(10)$ cm over a few-meter diameter].}  

The atmospheric neutrino fluxes $\Phi$ also depend, in general, on the three variables $(E',\,\theta',\,\phi')$.
Overall normalization errors are large [$O(10)\%$], but are also highly correlated between $(\mu,\,e)$ or $(\nu,\,\overline\nu)$, and can thus be reduced to a (few) percent by taking appropriate ratios. 
At this level of accuracy, a number of irreducible shape uncertainties are also known to be present. 
First-order shape variations have often been parametrized via ``tilts'' of both the energy spectrum index and the zenith-angle distributions (see, e.g., \cite{Tilt,PhDT}). This simplification is generally adequate when the investigated signal is sizable, as it happens for atmospheric $\nu_\mu$ disappearance due to the dominant $(|\Delta m^2|,\,\theta_{23})$ parameters, but may be too coarse for subdominant signals. A dedicated Workshop recognized, a decade ago, the need for a more refined characterization of atmospheric flux (and other) uncertainties, in view of future high-statistics experiments seeking subleading effects \cite{Kaji}. As a relevant follow-up, the authors of 
\cite{Barr} broke down the main atmospheric flux uncertainties into $N_S=26$ independent error sources within a one-dimensional propagation model for the Kamioka site, and studied their associated effects at the (few) percent level on the energy-angle spectra.%
\footnote{See
also \protect\cite{Hond} for an independent error assessment.} 
 It would be desirable to repeat this relevant study in a three-dimensional flux model for the South Pole site, possibly within two or more independent simulations, in order to identify an analogous set of $N_S$ systematics and associated flux variation functions 
$\delta_S\Phi^\alpha(E',\theta',\phi')$, to be included in data fits. At present, it is legitimate to
assume that the shapes of the $\Phi^\alpha(E',\,\theta',\,\phi')$ functions are not known to better than the (few) percent level.

Concerning the calculation of the oscillation probabilities $P_{\alpha\beta}$, the main uncertainties are related to variations of the mass-mixing neutrino parameters, particularly of the dominant ones $(|\Delta m^2|,\,\theta_{23})$; such variations are known to reduce the sensitivity to hierarchy effects \cite{Smir}. To a much lesser extent, the $P_{\alpha\beta}$ are affected (in the atmosphere) by 
uncertainties in the production height distribution, and (in matter) by uncertainties in the electron density profile.

The CC cross sections are poorly known in the (few) GeV range, especially when 
deep inelastic scattering is not dominant. Uncertainties on total and differential cross sections
affect the spectral normalization and shape, respectively, at a typical level of few \%. The shape uncertainties may affect  
high-statistics estimates of the dominant parameters $(\Delta m^2,\theta_{23})$ \cite{Cros}. A fortiori, such uncertainties cannot
be ignored in the extraction of subdominant effects. 

Detector response uncertainties add upon cross section errors in the determination of the energy-angle resolution functions, which characterize
the probability distributions  of the observable (reconstructed) kinematical parameters around the unobservable (true) ones.
The resolution functions are reported in \cite{PING} (see Fig.~8 therein) in terms of $E'$, but
they may also depend on $(\theta',\,\phi')$ via detector response anisotropies.  
In the absence of a calibration neutrino beam, the resolution functions can only be simulated, 
and their centroids and shapes are unavoidably affected by both cross section and 
reconstruction uncertainties.

Finally, the multi-dimensional integration into binned spectra may be a source of numerical errors in itself. 
For instance, in order to keep the calculations manageable, we have reduced the number of nested integrations, by averaging out {\em a priori\/} the azimuth angle, the production height, and the energy and angle(s) of the outgoing leptons (Sec.~II~D). We have also assumed gaussian resolution functions in order to apply, in each bin, the reduction in Eq.~(\ref{double}). Of course, these approximations can be avoided by a brute-force integration over all the relevant variables, including the (true and reconstructed) neutrino and lepton energies and directions. This approach (which is well beyond the scope of this paper) becomes impractical, however, when one must repeat the calculations by varying many systematics
in real or prospective data fits.
Note also that the oscillation probabilities may vary wildly over typical energy-angle bin ranges, making the calculations quite demanding in terms of integrand grid sampling and numerical accuracy. 

Alternatively, one might replace multi-dimensional integration into bins by a full Monte Carlo approach, randomly following the entire process of production, propagation, interaction, reconstruction and binning of events.%
\footnote{However, to our knowledge, there is no known MC code including, at the same time, both atmospheric neutrino production from cosmic rays and neutrino interactions in a South Pole detector, since the associated MCs have been developed by different groups of researchers.}
 In order to keep the statistical MC error at subpercent level, each bin should collect no less than $O(10^4)$ simulated events, which implies $O(10^{6-7})$ events for each MC spectrum with $O(10^2)$ bins. Moreover,     
the MC spectrum must be repeatedly calculated, by sampling the probability distributions of the floating variables in the fit (including the oscillation parameters and the systematics uncertainties), leading to no less than $O(10^{9})$ generated events, i.e., 
to a MC statistics several orders of magnitude higher than the real data sample of $O(10^5)$ events.  
It is not obvious (at least to us) that this goal, and the desirable subpercent statistical MC accuracy in each bin, can be 
practically reached. 
  We thus argue that any method of calculation of event rates can also 
induce (largely uncorrelated) residual errors in each bin, at about the percent level. These additional numerical
uncertainties may not be totally negligible in the statistical analysis of next-generation atmospheric neutrino experiments such as PINGU. 

In conclusion, we have discussed qualitative arguments in favor of many possible sources of (few) percent uncertainties on the shape of 
the energy-angle distributions in PINGU. Each of these sources gives extra freedom to adjust the event spectra in data fits, thus reducing
the sensitivity to spectral differences induced by hierarchy effects. Although each reduction may be small, 
their sum may become noticeable. In the following Section we shall study, in a more quantitative way,
the progressive impact of some of the above error sources in PINGU. 

\section{Statistical approach to the hierarchy sensitivity in PINGU}

In this Section we discuss our statistical approach to the hierarchy sensitivity in PINGU. 
We define the methodology used to deal with systematic errors, and discuss some subtle issues emerging in the limit of very high statistics. 
We then consider an increasingly large set of errors, including those coming from oscillation parameter and normalization uncertainties, 
from (known and unknown) correlated shape systematics, and from possible residual uncorrelated errors.   

\subsection{Methodology for systematics, and discussion of high-statistics limits}

Our statistical analysis of the $\mu$ and $e$ event spectra in PINGU is based on a $\chi^2$ approach. As argued in \cite{Sens}, a good metric for the sensitivity to the hierarchy (despite its discrete nature) is given by the marginalized $\Delta\chi^2$ difference between the true hierarchy (TH) and the wrong hierarchy (WH), 
where (TH,~WH) refer to either (NH,~IH) or (IH,~NH). The parameter $N_\sigma = \sqrt {\Delta\chi^2}$ can thus be taken as a shorthand for the 
effective number of standard deviations separating the TH and WH hypotheses.

The TH and WH spectral event rates are defined as
\begin{eqnarray}
R^\alpha_{ij}(p_k) &=& \frac{N^\alpha_{ij}(\text{TH};\,p_k)}{T}\ ,\\
\tilde R^\alpha_{ij}(\tilde p_k) &=& \frac{N^\alpha_{ij}(\text{WH};\,\tilde p_k)}{T}\ ,
\end{eqnarray}
where $T$ is the detector live time, the $p_k$ are the (oscillation and systematic) fixed parameters in TH, while $\tilde p_k$ are the corresponding floating
parameters in WH.%
\footnote{Since we neglect seasonal variations and take average fluxes from \cite{HKKM}, the event rates are constant in time.}
 The ``theoretical'' WH hypothesis is tested against the ``experimental data'' represented by the TH event rates, which are affected by statistical errors decreasing as $\sqrt{T}$,
\begin{equation}
s^\alpha_{ij} = \frac{\sqrt{R^\alpha_{ij}}}{\sqrt{T}} \ (1\sigma)\ ,
\end{equation}
and by systematic errors on the parameters $p_k$,
\begin{equation}
p_k \pm \sigma_k \ (1\sigma)\ .
\end{equation}

The $\Delta\chi^2$ function is defined as 
\begin{equation}
\Delta\chi^2 = \min_{\tilde p_k}\left[  \sum_{i=1}^{10} \sum_{j=1}^{16} \sum_{\alpha=\mu,\,e}\frac{\left(R^\alpha_{ij}(p_k)-\tilde R^\alpha_{ij}(\tilde p_k)\right)^2}{(s^\alpha_{ij})^2+(u^\alpha_{ij})^2} + \sum_k\left(\frac{p_k-\tilde p_k}{\sigma_k}\right)^2\right]\ ,
\label{chi2}
\end{equation}
where the second term represents the sum of penalty functions for the nuisance parameters $\tilde p_k$ (assuming gaussian errors $\sigma_k$). Special cases for the penalties are: (1) $\sigma_k=0$, which corresponds to having the $k$-th parameter fixed as $\tilde p_k= p_k$; and (2) $\sigma_k=\infty$, which corresponds to unconstrained values of $\tilde p_k$.  

In the above equation, the first term includes possible uncorrelated  errors $u^\alpha_{ij}$ which (as argued in Sec.~II~G) may stem, e.g., from finite MC statistics in each bin, or from residual systematics. There are however, deeper motivations to include such uncorrelated errors, as noted in
\cite{Lisi}. In fact,  the above $\Delta\chi^2$ function entails two strong assumptions: (1) that we know 
all the possible sources of correlated systematic parameters $\tilde p_k$; and (2) that we know exactly the effect of each $\tilde p_k$ variation 
on each binned rate 
$\tilde R^\alpha_{ij}$. In the limit of very large statistics ($s^\alpha_{ij}\to 0$), these two assumptions would 
lead to $\Delta\chi^2\to\infty$ for $u_{ij}=0$, 
unless $R^\alpha_{ij}(p_k)\equiv \tilde R^\alpha_{ij}(\tilde p_k)$ for some very peculiar values of $\tilde p_k$.
However, in general, no combination of $\tilde p_k$ can exactly reproduce a generic data set $R^\alpha_{ij}$. 
In other words, the assumptions that we do know all
the systematics and all their spectral effects leads to the paradoxical conclusion that the sensitivity $N_\sigma =\sqrt{\Delta\chi^2}$  
grows indefinitely with $\sqrt{T}$, without reaching any reasonable, systematics-limited plateau \cite{Lisi}. 

This situation is basically unprecedented in neutrino experiments, usually characterized by relatively low statistics.%
\footnote{Actually, current short-baseline reactor experiments with huge statistics (million events) are now facing
such challenges in the characterization of spectral shape systematics, see \protect\cite{Dwye} and references therein.}      
 To solve the paradox, one should admit that the knowledge of spectral systematics may be either incomplete or inaccurate to some extent, and try to deal with
 the residual ignorance.  One possibility is to render the spectra more flexible, by including additional families of admissible spectral deviations via extra parameters $\tilde p_k$, constrained by dedicated considerations or educated guesses. This method has been explored, e.g., in the context of fits to parton distribution functions \cite{Pump,Ball} and to  precision cosmological data \cite{Kitc,Tayl}.%
\footnote{Pushing this approach further, the likelihood could eventually be minimized over a functional ensemble (via path integral techniques) rather than over a discrete
$\tilde p_k$ ensemble, see \protect\cite{Path}.}
 Another possibility is to parametrize our ignorance by allowing additional uncorrelated errors $u_{ij}$ of reasonable size in each bin, which lead to finite
 $\Delta\chi^2$ values in the limit of 
 infinite statistics \cite{Lisi}. In our analysis, we shall study in sequence the effects of both approaches (which, in a sense, try to deal with ``errors on errors''  \cite{Erro}).

The $\chi^2$ minimization procedure is also nontrivial for large numbers of bins $i\times j$ and 
of floating parameters $\tilde p_k$. In the context of PINGU, a brute-force scan of the $\{\tilde p_k\}$ parameter space through a fixed sampling grid would be prohibitively time consuming, also because the hierarchy sensitivity can be as large as $N_\sigma\sim O(10)$ in favorable cases, implying that the 
$\tilde p_k$ distribution tails must be sampled at the same level. Alternatively, one might 
explore the large $\{\tilde p_k\}$ parameter space via a Markov Chain Montecarlo (MCMC) method \cite{MCMC}.
This method (as the brute-force scan) does not make any hypothesis on the functional form of the functions 
$\tilde R^\alpha_{ij}(\tilde p_k)$, and can work also for generic
(non-quadratic) penalty functions for the $\tilde p_k$. However, 
we have found that, at least in our implementation of the MCMC \cite{MCMC,Capo} 
for the PINGU analysis, the cases with high sensitivity $N_\sigma \sim O(10)$ are numerically too demanding 
and time consuming; therefore, we have abandoned this approach.
We have finally chosen to use a method mainly based on the  ``pull approach'' \cite{Pull}, which increases enormously the minimization speed and stability, at the price of
of assuming a first-order expansion of the $\tilde R^\alpha_{ij}(\tilde p_k)$ under small deviations of (some) parameters $\tilde p_k$ around 
the ``true'' values $p_k$: 
\begin{equation}
\tilde p_k = p_k + \xi_k \ \longrightarrow \ \tilde R_{ij}^\alpha(\tilde p_k) \simeq \tilde R_{ij}^\alpha(p_k) + \xi_k \left(
\frac{\partial \tilde R_{ij}^\alpha(\tilde p_k)}{\partial \tilde p_k}\right)_{\tilde p_k=p_k}\ .
\label{Pulls}
\end{equation}

In the context of our PINGU analysis, the linearization of the rates
is actually exact for some parameters (e.g., normalization errors), and is a
reasonably good approximation for all the other parameters, with the only exception of the mixing angle $\theta_{23}$ and the CP-violating phase $\delta$,
whose dependence cannot be linearized at all. For this reason, for any given choice of TH parameters $(\sin^2\theta_{23},\delta)$, 
we do scan the WH parameters $(\sin^2 \tilde \theta_{23},\tilde \delta)$ over a grid sampling the full range $[0,\,1]\otimes [0,\,2\pi]$.
For each $(\sin^2 \tilde \theta_{23},\tilde \delta)$ point of the the grid, we numerically calculate the derivatives in Eq.~(\ref{Pulls})
by taking finite differences at $\pm2\sigma_k$, then minimize the $\chi^2$ analytically over the linear(ized) $\tilde p_k$ variations \cite{Pull}, and finally find the absolute minimum by scanning the whole 
$(\sin^2 \tilde \theta_{23},\tilde \delta)$ grid.

\subsection{Systematics due to oscillation and normalization uncertainties}

The most obvious sources of systematic errors are due to: (1) the calculation of oscillation probabilities and (2) absolute and relative normalizations.
Concerning the first, we attach the following $1\sigma$ fractional uncertainties to the central values in Eqs.~(\ref{Dm2}) and (\ref{q13}) \cite{Glob1}:
\begin{eqnarray}
\sigma(\Delta m^2)&=&2.6\%\ ,\\
\sigma(\sin^2\theta_{13})&=&  8.5\% \ . 
\end{eqnarray}
The parameters $\delta m^2$ and $\sin^2\theta_{12}$ are kept fixed as in Eqs.~(\ref{dm2}) and (\ref{q12}), respectively, since their
errors induce negligible effects in the PINGU analysis. As already discussed, the true parameter $\delta$ is fixed at $3\pi/2$ as in Eq.~(\ref{dcp}), while 
for the wrong hierarchy it is left free in the range $[0,\,2\pi]$. The true parameter $\sin^2\theta_{23}$ is chosen in the range $[0.4,\,0.6]$, while
for the wrong hierarchy it is left free in the range $[0,\,1]$.%
 Finally, we add a reasonable $3\%$ error on the electron density in the Earth's core,
\begin{equation}
\sigma (N_e) =  3\%\ (\text{core})\ ,
\end{equation} 
to account for uncertainties in its chemical composition.%
\footnote{The typical difference between the $N_e$ values in the mantle and in the core is $\sim 6\%$ \protect\cite{Eart}.}

Concerning the absolute normalization, we attach an overall 15\% error $f_N$ to all the event rates, accounting for fiducial volume, flux and cross section uncertainties,
\begin{equation}
\tilde R^\alpha_{ij} \to \tilde R^\alpha_{ij}(1+f_N)\ , \ \sigma(f_N)= 0.15\ .
\end{equation}
The relative normalizations between the $\mu$ and $e$ rates, and between the $\nu$ and $\overline\nu$ components of the rates,
are allowed to differ, respectively, up to $ 8\%$ and $ 6\%$ at $1\sigma$  (which represent values in typical ranges \cite{Wint,PhDT})%
\footnote{In principle, due to oscillations, one should distinguish between a relative $\Phi^\mu/\Phi^e$ flux error at the source, and a relative $\tilde R^\mu_{ij}/\tilde R^e_{ij}$
mis-identification error at the detector. However, we have verified that these errors are 
highly correlated in the fit results, and their merging in a single error source is a justified approximation within the present work.}
\begin{equation}
\left(
\begin{array}{cc}
\tilde R^\mu_{ij}\\
\tilde R^e_{ij}
\end{array}
\right)\to
\left(
\begin{array}{cc}
\tilde R^\mu_{ij}(1+\frac{1}{2}f_R)\\
\tilde R^e_{ij}(1-\frac{1}{2}f_R)
\end{array}
\right)\ , \ \sigma(f_R)= 0.08\ ,
\end{equation}
\begin{equation}
\left(
\begin{array}{cc}
\tilde R^\alpha_{ij}(\nu)\\
\tilde R^\alpha_{ij}(\overline\nu)
\end{array}
\right)\to
\left(
\begin{array}{cc}
\tilde R^\alpha_{ij}(\nu) (1+\frac{1}{2}f_\nu)\\
\tilde R^\alpha_{ij}(\overline\nu) (1-\frac{1}{2}f_\nu)
\end{array}
\right)\ , \ \sigma(f_\nu)= 0.06\ .
\end{equation}
Note that, at this stage, we are not consider further systematics, either correlated ($\tilde p_k$) or uncorrelated ($u^\alpha_{ij}$), which could 
give more ``flexibility'' to the spectral shape.

Let us discuss two examples of best-fit spectra in a ``favorable'' and ``unfavorable'' case for PINGU, including the above uncertainties.
In general,  the case of true normal hierarchy is more favorable, since the corresponding
matter effects are stronger for neutrinos, which are enhanced by a larger cross section than antineutrinos. Cases  
with $\sin^2\theta_{23}$ in the second octant are also more favorable, since this parameter 
modulates the amplitude of the relevant
oscillation channel $P_{\mu e}$, which embeds large matter effects.  
Conversely, cases with true inverted hierarchy and 
$\sin^2\theta_{23}$ in the first octant are typically less favorable for hierarchy discrimination.

Figure~3 shows the absolute spectra $R^\mu_{ij}(p_k)$ and $R^e_{ij}(p_k)$ in terms of events per bin (upper and lower left panels, respectively) for the
favorable case of true normal hierarchy and $\sin^2\theta_{23}=0.6$. The middle panels show the statistical differences between such spectra and the 
corresponding best-fit spectra in inverted hierarchy, $\tilde R^\mu_{ij}(\tilde p_k)$ and $\tilde R^e_{ij}(\tilde p_k)$, after marginalization over
the $\tilde p_k$ systematic parameters (oscillation and normalization uncertainties) described above. As pointed out in several papers 
\cite{Smir,PING,Fran,Wint,Blen,Colo,Hagi}, statistical differences, up to 1--2$\sigma$ in some bins, 
appear in the energy-angle region where matter effects are generically large, and even beyond (due to smearing); 
the differences typically change sign by changing flavor and, for a given flavor, they also change sign in the energy-angle plane. 
Such patterns of statistical deviations should thus provide useful cross-checks, provided that they are not spoiled by systematic effects.
The right panels show the same differences, expressed in terms of percent deviations, reaching a few~\% for muon event spectra and 
twice as much for electron event spectra. Hypothetical systematic shape deviations at the 5--10\% level, ``equal and opposite'' to those shown in 
the right panels of Fig.~3, would basically cancel the hierarchy difference, strongly reducing the related PINGU sensitivity. A nonegligible reduction
can still be expected, however, for smaller shape deviations at the (few) percent level.

\newpage
\phantom{.}
\begin{figure}[t]
\centering
\includegraphics[height=8.6cm]{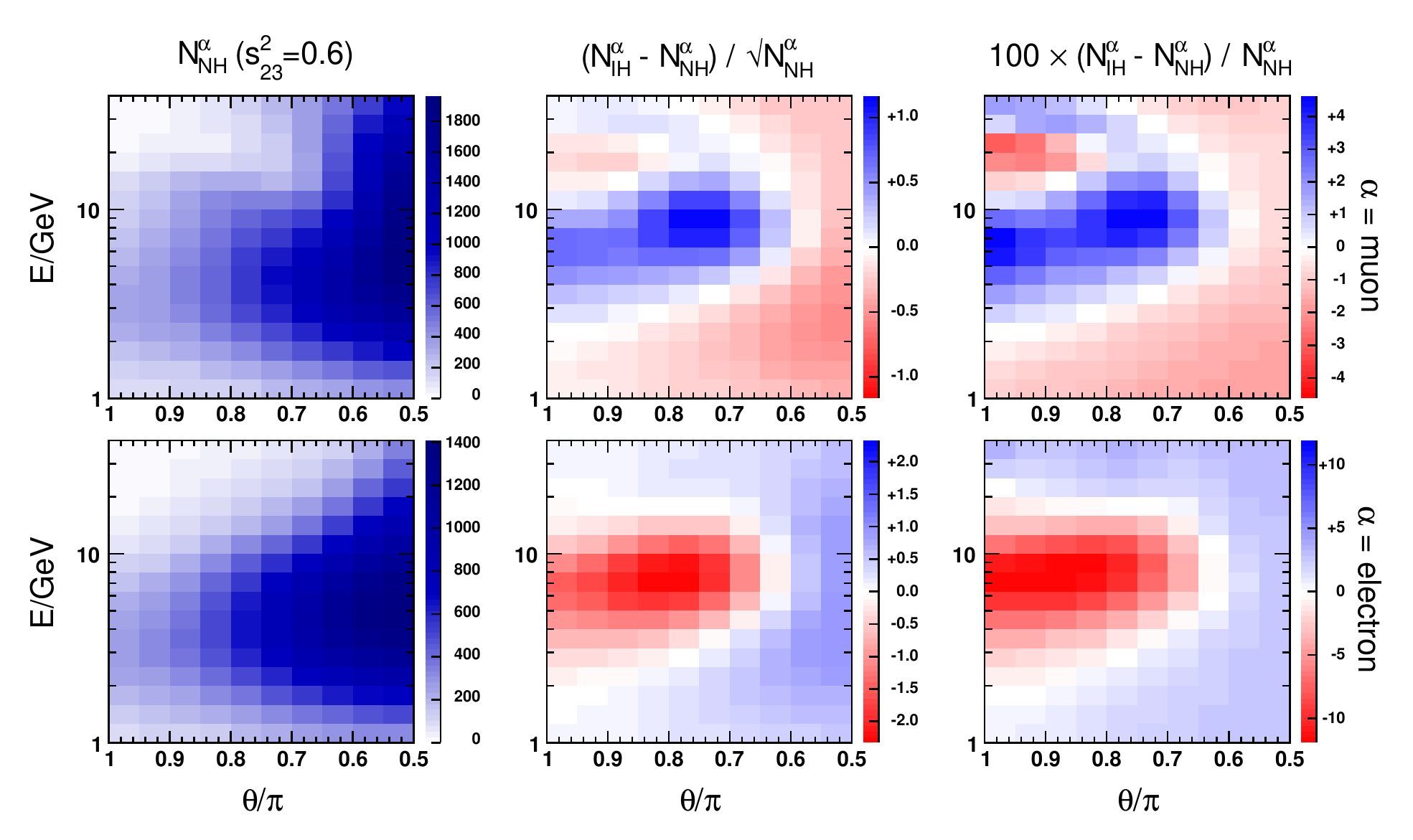}
\caption{Case of true normal hierarchy and $\sin^2\theta_{23}=0.6$. Left panels: absolute event spectra (top: $\mu$ events; bottom: $e$ events). 
Middle panels: statistical deviations with respects to the best-fit spectrum in the wrong (inverted) hierarchy, marginalized over oscillation and normalization systematics only. Right panels: the same deviations in percent values.
\label{Fig_03}}
\end{figure}
\phantom{.}
\begin{figure}[h]
\centering
\includegraphics[height=8.6cm]{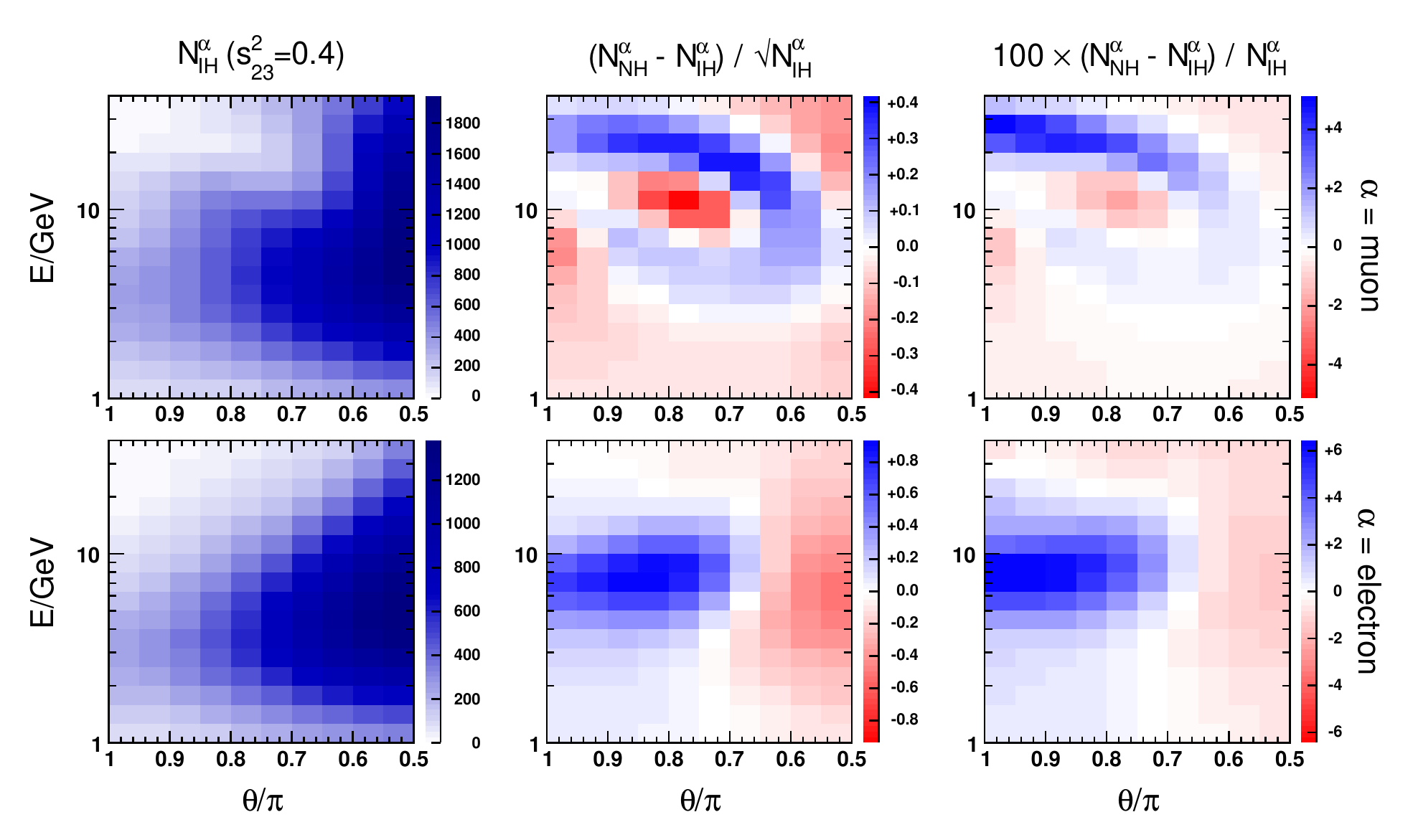}
\caption{As in Fig.~3, but for true inverted hierarchy and $\sin^2\theta_{23}=0.4$.
\label{Fig_04}}
\end{figure}
\phantom{.}
\newpage

Figure~4 is analogous to Fig.~3, but is obtained for the ``unfavorable'' case of true inverted hierarchy and for $\sin^2\theta_{23}=0.4$. In comparison 
with Fig.~3, the results in Fig.~4 show the following features:
(1) the left panels are basically
indistinguishable, confirming that the hierarchy discrimination can only emerge from careful spectral analyses and not ``by eye;''
(2) the deviations in the middle and right panels are, as expected, generally smaller (and with opposite sign) 
with respect to those in Fig.~3, making systematic shape deviations at the few~\% level more dangerous 
for hierarchy discrimination; and (3) the detailed patterns of deviations in Fig.~4 
are somewhat different from Fig.~3, especially for $\mu$ spectra, where the deviations can
 extend to relatively high energies and change sign twice; these features may be traced to the fact that the wrong (NH)
 $\mu$ spectrum tries to fit the true (IH) spectrum also via deviations of the dominant oscillation parameters, 
which induce a mismatch in the region of the first minimum for $\nu_\mu$ disappearance.

In conclusion,  Figs.~3 and 4 show, once more, the necessity to investigate the impact of {\em shape\/} systematics at the (few) percent level
in a wide energy-angle region, not necessarily restricted to nearly upgoing events at few~GeV. For this purpose, following
the discussion in Sec.~II~G, we shall first include ``known'' systematics directly affecting the shape, 
and then try to parametrize ``unknown''  uncertainties.

\subsection{Adding energy-scale and resolution width uncertainties}

Uncertainties in the double differential cross-section, as well as in the detector energy-angle reconstruction, eventually affect
the shapes of the resolution functions. For the sake of simplicity, we consider only gaussian resolution functions
[see Eqs.~(\ref{resole}) and (\ref{resolq})], which may be affected by two kinds of systematics: biases in the centroid, and fluctuations 
in the width. More complicated (e.g., skewed) variants might be considered for nongaussian cases. 

Following recent PINGU presentations (see, e.g., \cite{Dawn}), we assume 
that the true energy centroid may be biased by up to 5\% at $1\sigma$,
\begin{equation}
E' \to E' (1+f_E)\ , \ \sigma(f_E)=0.05\ .
\end{equation}
We actually include two such energy scale errors, for $\mu$ and $e$ event independently, while we neglect possible
directional biases for such events, which are expected to be smaller (and are usually undeclared in PINGU presentations).

Concerning the resolution widths, on the basis of
our histogram fitting procedure described in Sec.~II~C, as 
well as on fluctuations in the PINGU own evaluation of widths \cite{Dawn}, we estimate that they may vary up to 10\% at $1\sigma$,
independently of each other:
\begin{eqnarray}
r^\alpha_z \to r^\alpha_z(1+f^\alpha_z)\ , \ \sigma(f^\alpha_z)=0.1\ ,
\end{eqnarray}
where $\alpha=\mu,\,e$ and $z=E,\,\theta$. The allowance for  slightly wider or narrower resolution functions
is a relevant degree of freedom in the fit, given the  role of smearing effects in determining the observable spectral shapes.

\subsection{Parametrizing residual correlated systematics with polynomials}

The previous energy scale and resolution width errors do not exhaust the (presumably long) list of shape systematics. 
As argued in Sec.~II~G, uncertainties in the effective volume and in the reference atmospheric fluxes 
may also lead to an entire set of (few) percent deviations as a function of energy and angle,
which are not necessarily well known or under good control. Usually, this ignorance is parametrized in terms of first-order deviations 
(``tilts'') of the spectra but, in the context of PINGU, one should allow for further (smooth) nonlinear deviations,
which may be more crucial for hierarchy discrimination, as shown by Figs.~3 and 4. Nonlinear systematics at 
(few) percent level are known to affect, e.g., atmospheric neutrino flux shapes in both energy and angle \cite{Barr},
as discussed in Sec.~II~G.

In the absence of a detailed study of such residual shape systematics in PINGU, we provisionally assume that
the observable PINGU spectra have small, additional functional uncertainties, parametrized in terms of polynomials. 
In particular, we rescale the abscissa and ordinate variables of the spectra in Figs.~1--4 as 
\begin{eqnarray}
x &=& 4\theta/\pi - 3 \ , \\
y &=& 1.25 \log_{10}(E/\text{GeV}) - 1\ ,
\end{eqnarray}
so that they range within
\begin{equation}
(x,\,y) \in [-1,\,+1]\otimes [-1,\,+1]\ .
\end{equation}

We assume that the rates $\tilde R^\alpha_{ij}$ may be subject to generic  $h$-degree polynomial deviations in $(x,\,y)$ of the kind:
\begin{equation}
\tilde R^\alpha_{ij} \to \tilde R^\alpha_{ij}\left(1+\sum_{m+n>0}^{h} c_{nm}^\alpha x^m_i y^n_j \right)\ ,
\label{poly}
\end{equation}
where $x_i$ and $y_j$ are the midpoint coordinates of the $ij$-th bin. The coefficients $c^\alpha_{nm}$ are allowed to float around a null central value
within representative errors, that we choose as
\begin{equation}
\sigma(c^\alpha_{nm})=10^{-2} \times \left\{
\begin{array}{ll}
1.5& \text{(default)}\ ,\\
3.0& \text{(doubled\ errors)}\ ,\\
.75& \text{(halved\  errors)}\ ,
\end{array}
\right.
\label{case}
\end{equation}
so as to cover cases with shape systematics at various levels (percent, few percent, subpercent).

We shall consider polynomials
with $h=1$, 2, 3 and 4, corresponding to linear, quadratic, cubic, and quartic deformations in $x$ and/or $y$. 
Such polynomial shape deformations are easily implemented in the statistical analysis, 
since the rates $R^\alpha_{ij}$ are linear in the additional pull parameters $c^\alpha_{nm}$,
which provide from 4 (linear case) to 28 (quartic case) extra degrees of freedom in the fit.%
\footnote{  
 Note that, in Eq.~(\ref{poly}), the terms $c^\alpha_{00}$ are omitted,
since they correspond to the overall normalization errors of Sec.~III~B. Note also that the linear terms $c^\alpha_{10}$ and $c^\alpha_{01}$
parametrize the usual spectral ``tilts'' discussed in Sec.~II~G.}  
By construction, the $\pm1\sigma$ excursion of a generic term $c^\alpha_{nm}x_i^n y_j^m$ is thus
contained within $\pm 1.5\%$ in the default case [and, similarly, within $\pm 3\%$ and $\pm0.75\%$ in the other two cases of Eq.~(\ref{case})]. 
Although several terms $c^\alpha_{nm}x_i^n y_j^m$ might add up to much more than $\pm 1.5\%$ in the default case, such a freedom is never
really exploited in the fit, leaving the best-fit spectral deformation at a relatively small, few-percent level, as we shall comment in the next Section~IV.

\subsection{Parametrizing residual uncorrelated systematics}

It is legitimate to posit (as argued in Sec.~II~G) that our knowledge of the systematics, no matter how detailed, may be incomplete, leaving 
residual uncorrelated errors $u^\alpha_{ij}$ in each bin from various sources (including finite statistics effects in atmospheric, cross section, and
reconstruction MC simulations). We shall consider three simple, representative cases for the residual (uncorrelated) fractional
uncertainties in each bin, namely:   
\begin{equation}
\frac{u^\alpha_{ij}}{R^\alpha_{ij} }=10^{-2} \times \left\{
\begin{array}{ll}
1.5& \text{(default)}\ ,\\
3.0& \text{(doubled\ errors)}\ ,\\
.75& \text{(halved\ errors)}\ .
\end{array}
\right.
\label{case2}
\end{equation}
This completes our list of uncertainties, which will be progressively included, via Eqs.~(\ref{chi2}) and (\ref{Pulls}), in the following statistical analysis.

\section{Statistical analysis of the hierarchy sensitivity: results}

Figure~5 shows the PINGU sensitivity to the hierarchy, in terms of standard deviations separating the true mass hierarchy (top: NH; bottom: IH)
from the wrong mass hierarchy, as a function of
the detector live time $T$ in years. The bands cover the fit results obtained by spanning the range $\sin^2\theta_{23}|_\text{true} \in[0.4,\,0.6]$.
The abscissa is scaled as $\sqrt{T}$, so that the  bands would grow linearly in the ideal case of no systematic errors 
(not shown). 
From left to right,  the fit includes the following
systematic errors: oscillation and normalization uncertainties, energy scale and resolution width errors, polynomial shape systematics
(with up to quartic terms), and uncorrelated systematics, as defined in  Sec.~III. The last two error sources are kept at the default level of 1.5\%. 
With only normalization and systematic errors, $N_\sigma$ grows almost linearly in $\sqrt{T}$, i.e., the experiment is not limited by these systematics, even after 10
years of data taking. However, the progressive inclusion of correlated shape systematics, both ``known'' (resolution scale and widths) and ``unknown'' 
(ad hoc polynomial deviations), and eventually of uncorrelated shape systematics, provide a suppression of $N_\sigma$, whose estimated ranges
increase more slowly than $\sqrt{T}$. The typical effect of all the systematic shape errors in the rightmost panels
is to decrease 
the 5-year (10-year) PINGU sensitivity by up to $\sim 35\%$ ($\sim 40\%$), 
with respect to the leftmost panels in  Fig.~5. 
 
\begin{figure}[t]
\includegraphics[height=10cm]{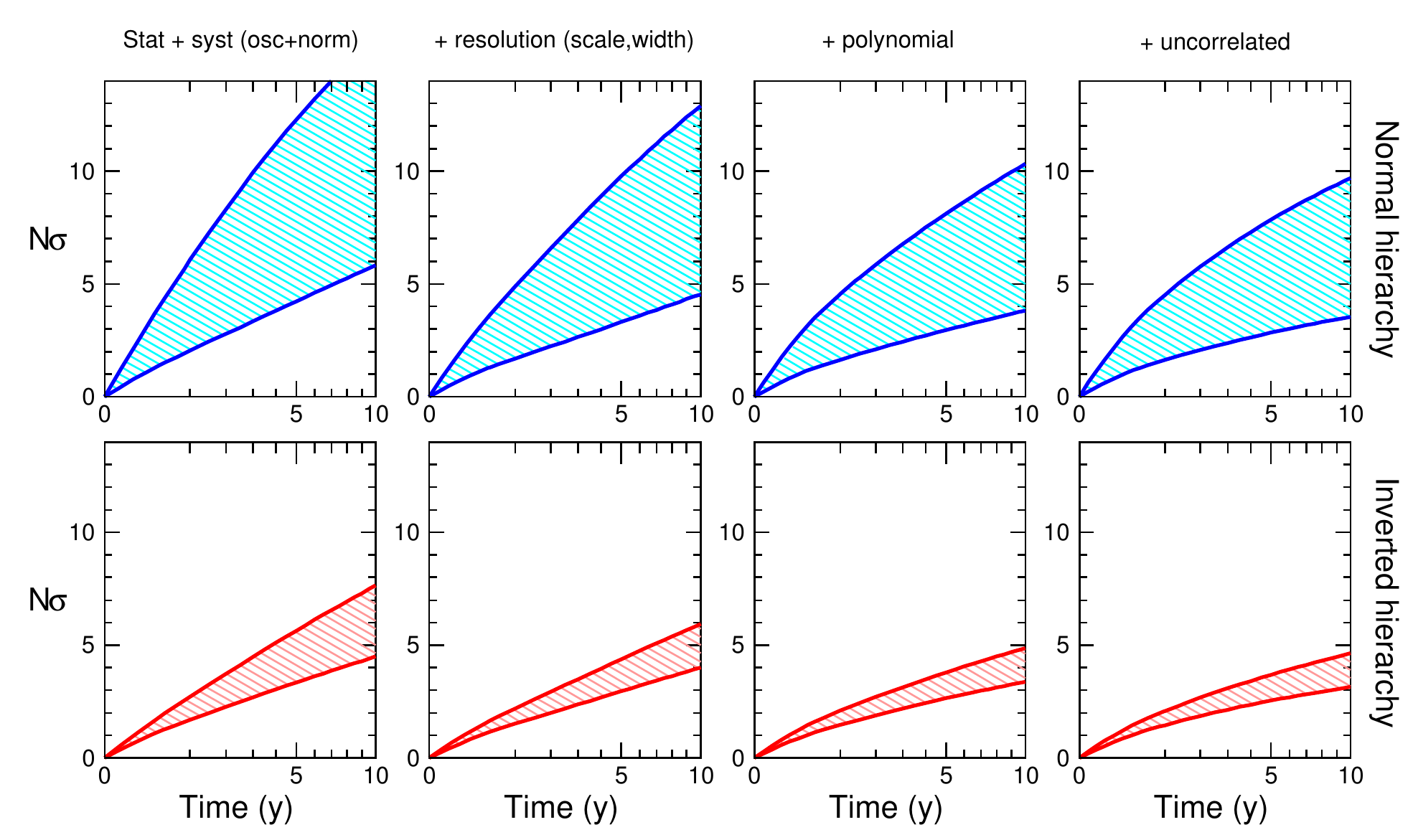}
\caption{PINGU sensitivity to the hierarchy ($N_\sigma$), 
for either true NH (top panels) or true IH (bottom panels), as a function of the live time $T$ in years. The abscissa is scaled as $\sqrt{T}$, so that 
the sensitivity bands (which span $\sin^2\theta_{23}|_\text{true}\in [0.4,\,0.6]$)  would grow linearly 
for purely statistical errors. From left to right, the fit includes the following
systematic errors: oscillation and normalization uncertainties, energy scale and resolution width errors, polynomial shape systematics
(with up to quartic terms) at the 1.5\% level, and uncorrelated systematics at the 1.5\% level, as defined in  Sec.~III. 
\label{Fig_05}}
\vspace*{-0mm}
\end{figure}

Table~I reports numerical results for the same fit of Fig.~5, with a breakdown of the polynomial shape systematics 
(from linear to quartic deviations). It can be seen that most of the sensitivity reduction due to polynomial shape variations
is already captured at the level of linear and quadratic parametrization, with higher-degree terms contributing a small fraction of
$1\sigma$. Although each polynomial term $c^\alpha_{nm}x_i^n y_j^m$ can typically contribute to a $\pm1.5\%$ deviation by construction,
their sum $\sum c^\alpha_{nm}x_i^n y_j^m$ yields typical deviations of about $\pm 1\%$ ($\pm 2\%$) for $\mu$ ($e$) events in the fit, 
except for the case of NH in the second octant, where they can become twice as large (but where $N_\sigma$ is also large). 
In conclusion, reasonable shape uncertainties at the (few) percent level may produce a noticeable overall effect
on the PINGU sensitivity, although none of them appears to be crucial in itself.

\begin{table}[h]
\caption{Reduction of the PINGU sensitivity to the hierarchy (expressed in terms of $N_\sigma$ range for $\sin^2\theta_{23}\in [0.4,\,0.6]$) 
due to the progressive inclusion of various shape systematics, for 5 and 10 years of exposure. Correlated polynomial and uncorrelated 
systematic uncertainties are taken at the default level of $1.5\%$. See the text for details.}
\centering 
\begin{ruledtabular}
\begin{tabular}{lcccc}
& \multicolumn{2}{c}{5-year sensitivity $N_\sigma$}& \multicolumn{2}{c}{10-year
sensitivity $N_\sigma$}\\
 Errors included in the fit & True NH & True IH & True NH & True IH \\
 \hline
Stat. + syst (osc.+norm.)	& 4.23--12.3 & 3.34--5.64 & 5.82--16.1 & 4.49--7.64 \\
+ resolution (scale, width) & 3.31--9.76 & 2.95--4.37 & 4.54--12.9 & 4.00--5.94 \\
+ polynomial (linear) 		& 3.14--9.17 & 2.86--4.16 & 4.23--11.9 & 3.81--5.49 \\
+ polynomial (quadratic) 	& 3.01--8.29 & 2.69--3.88 & 3.93--10.6 & 3.47--5.05 \\
+ polynomial (cubic) 		& 2.98--8.26 & 2.67- 3.84 & 3.87--10.5 & 3.42--4.94 \\
+ polynomial (quartic) 		& 2.95--8.12 & 2.64--3.79 & 3.82--10.3 & 3.37--4.87 \\
+ uncorrelated systematics 	& 2.84--7.84 & 2.54--3.68 & 3.55--9.69 & 3.14--4.63 \\
\hline
Total $N_\sigma$ reduction 
from 1st row  				& 33--36\%	 & 24--35\%   & 39--40\%   & 30--39\%
\label{syst1}
\end{tabular}
\end{ruledtabular}
\end{table} 

\begin{figure}[t]
\includegraphics[height=9.1cm]{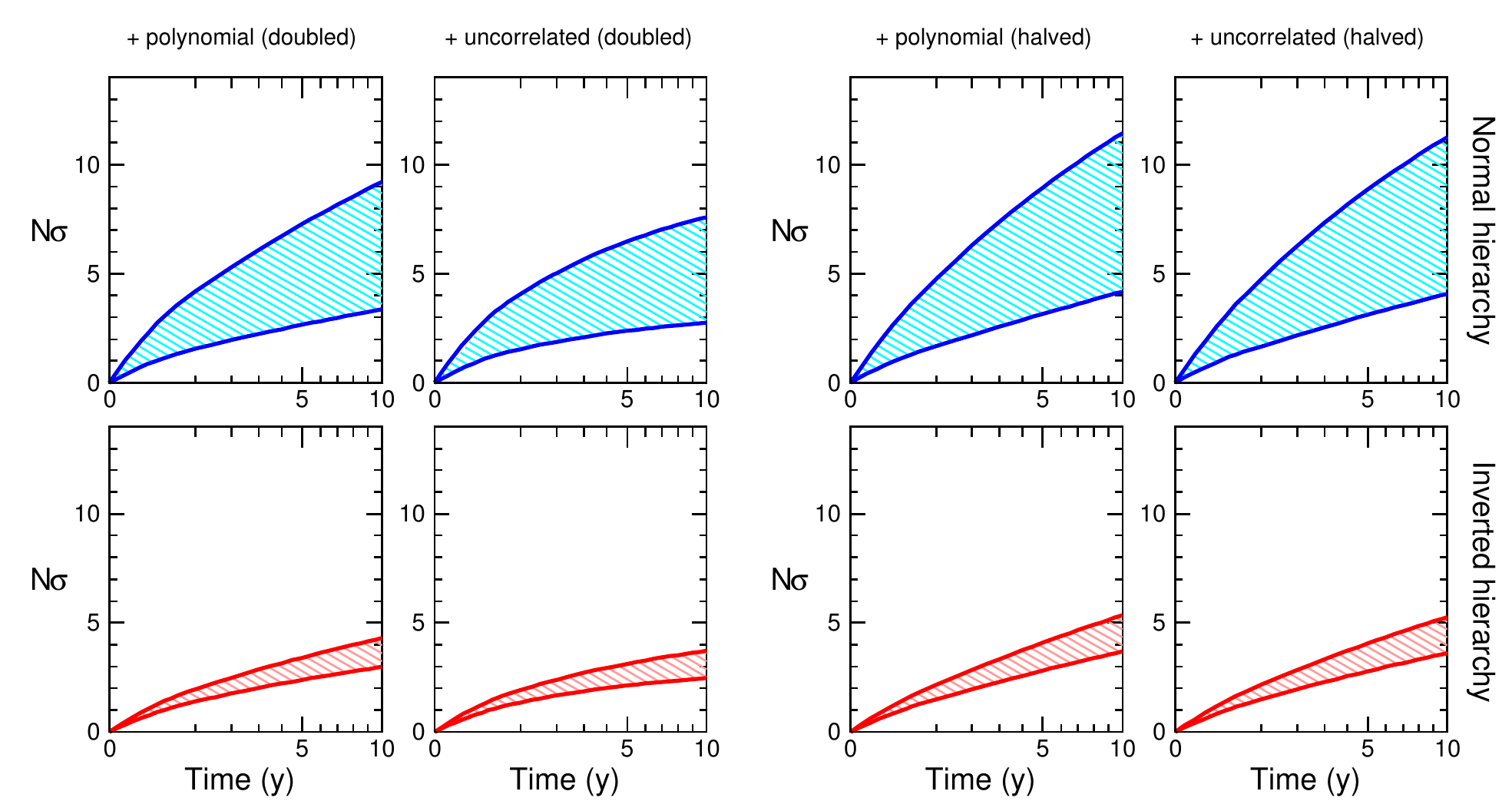}
\caption{As in Fig.~5, but for polynomial and uncorrelated systematic errors doubled  
(left) or halved (right), while statistical errors and systematics related to 
oscillation, normalization and resolution uncertainties are assumed to be the same. 
\label{Fig_06}}
\vspace*{-0mm}
\end{figure}

Figure~6 reports results analogous to the second half of Fig.~5, but with 
both the correlated polynomial and uncorrelated 
systematic uncertainties doubled, at the level of $3\%$ (left) or halved, at the level
of $0.75\%$ (right). The left panels show that, with shape systematics at the
few percent level, the hierarchy sensitivity tend to saturate in time, and can be  
lower than $\sim 3\sigma$ in the worst cases, even after 10 years of data taking.
Conversely, the right panels show that, with subpercent shape systematics, the 
sensitivity remains safely above $3\sigma$ (after 10 years) in all cases.  

Table~II is analogous to Table~I, but refers to the case where correlated polynomial and uncorrelated 
systematic uncertainties are doubled,
as in Fig.~6 (left). By construction,
the first two numerical rows are identical to Table~I, while the others show a more
pronounced reduction of the PINGU sensitivity, up to a factor of two when
all the errors are included.

\begin{table}[b]
\caption{As in Table~I, but with correlated polynomial and uncorrelated 
systematic uncertainties taken at the level of $3\%$. }
\centering 
\begin{ruledtabular}
\begin{tabular}{lcccc}
& \multicolumn{2}{c}{5-year sensitivity $N_\sigma$}& \multicolumn{2}{c}{10-year
sensitivity $N_\sigma$}\\
 Errors included in the fit & True NH & True IH & True NH & True IH \\
 \hline
Stat. + syst (osc.+norm.)	& 4.23--12.3 & 3.34--5.64 & 5.82--16.1 & 4.49--7.64 \\
+ resolution (scale, width) & 3.31--9.76 & 2.95--4.37 & 4.54--12.9 & 4.00--5.94 \\
+ polynomial (linear) 		& 3.03--8.79 & 2.77--3.94 & 4.07--11.5 & 3.68--5.16 \\
+ polynomial (quadratic) 	& 2.77--7.48 & 2.46--3.60 & 3.58--9.72 & 3.13--4.72 \\
+ polynomial (cubic) 		& 2.70--7.34 & 2.41- 3.46 & 3.43--9.32 & 3.02--4.40 \\
+ polynomial (quartic) 		& 2.66--7.27 & 2.38--3.40 & 3.36--9.21 & 2.98--4.29 \\
+ uncorrelated systematics 	& 2.37--6.48 & 2.12--3.12 & 2.75--7.50 & 2.47--3.72 \\
\hline
Total $N_\sigma$ reduction 
from 1st row  				& 33--53\%	 & 37--45\%   & 46--47\%   & 45--49\%
\label{syst2}
\end{tabular}
\end{ruledtabular}
\end{table} 

Finally, Table~III refers to the case where correlated polynomial and uncorrelated 
systematic uncertainties are halved,
as in Fig.~6 (right). In this case, such 
uncertainties do not play a relevant role, and the reduction of the sensitivity 
(up to about 20--30\%) is mainly due to resolution systematics.

In conclusion, the results of Figs.~5 and 6 and of Tables~I--III suggest that spectral shape uncertainties
require a careful investigation,  since they may be able to 
lower the PINGU sensitivity from 20\% to 50\%, as compared with an analysis 
including only the most obvious systematics due to oscillation and normalization
uncertainties.

\begin{table}[t]
\caption{As in Table~I, but with correlated polynomial and uncorrelated 
systematic uncertainties taken at the level of $0.75\%$. }
\centering 
\begin{ruledtabular}
\begin{tabular}{lcccc}
& \multicolumn{2}{c}{5-year sensitivity $N_\sigma$}& \multicolumn{2}{c}{10-year
sensitivity $N_\sigma$}\\
 Errors included in the fit & True NH & True IH & True NH & True IH \\
 \hline
Stat. + syst (osc.+norm.)	& 4.23--12.3 & 3.34--5.64 & 5.82--16.1 & 4.49--7.64 \\
+ resolution (scale, width) & 3.31--9.76 & 2.95--4.37 & 4.54--12.9 & 4.00--5.94 \\
+ polynomial (linear) 		& 3.24--9.51 & 2.92--4.30 & 4.39--12.4 & 3.92--5.77 \\
+ polynomial (quadratic) 	& 3.19--9.12 & 2.84--4.16 & 4.25--11.7 & 3.74--5.47 \\
+ polynomial (cubic) 		& 2.18--9.07 & 2.84- 4.14 & 4.22--11.6 & 3.73--5.43 \\
+ polynomial (quartic) 		& 3.15--8.94 & 2.81--4.09 & 4.17--11.5 & 3.69--5.34 \\
+ uncorrelated systematics 	& 3.11--8.86 & 2.78--4.06 & 4.08--11.3 & 3.61--5.26 \\
\hline
Total $N_\sigma$ reduction 
from 1st row  				& 26--28\%	 & 17--28\%   & 29-30\%   & 20--31\%
\label{syst3}
\end{tabular}
\end{ruledtabular}
\end{table} 

\section{Interplay between hierarchy and $\sin^2\theta_{23}$ determinations}

In the previous Section, we have assumed a prior range $\sin^2\theta_{23}|_\text{true} \in[0.4,\,0.6]$,
roughly corresponding to the current $\pm2\sigma$ allowed region \cite{Glob1}.
In view of future constraints on $\theta_{23}$ coming from ongoing and future accelerator experiments, it
is useful to consider also a possible reduction of this range in prospective PINGU analyses. 

In particular, let us consider Fig.~7, which is analogous to Fig.~5, but is obtained for
$\sin^2\theta_{23}|_\text{true} \in[0.46,\,0.54]$. One can notice a 
significant narrowing of the $N_\sigma$ bands, and an overall gain in 
the minimum sensitivity. However, the pattern of progressive reduction of $N_\sigma$
due to the inclusion of various shape systematics is similar to the one in Fig.~5.
Therefore, prior information on $\sin^2\theta_{23}$
is of crucial relevance in determining the {\em absolute\/} sensitivity
to the hierarchy, although it does not affect the {\em relative\/} reduction effects
of systematic shape uncertainties. In this context, it
makes sense to study how well this mixing angle may be determined by PINGU itself.

\vspace*{1cm}
\begin{figure}[b]
\includegraphics[height=10cm]{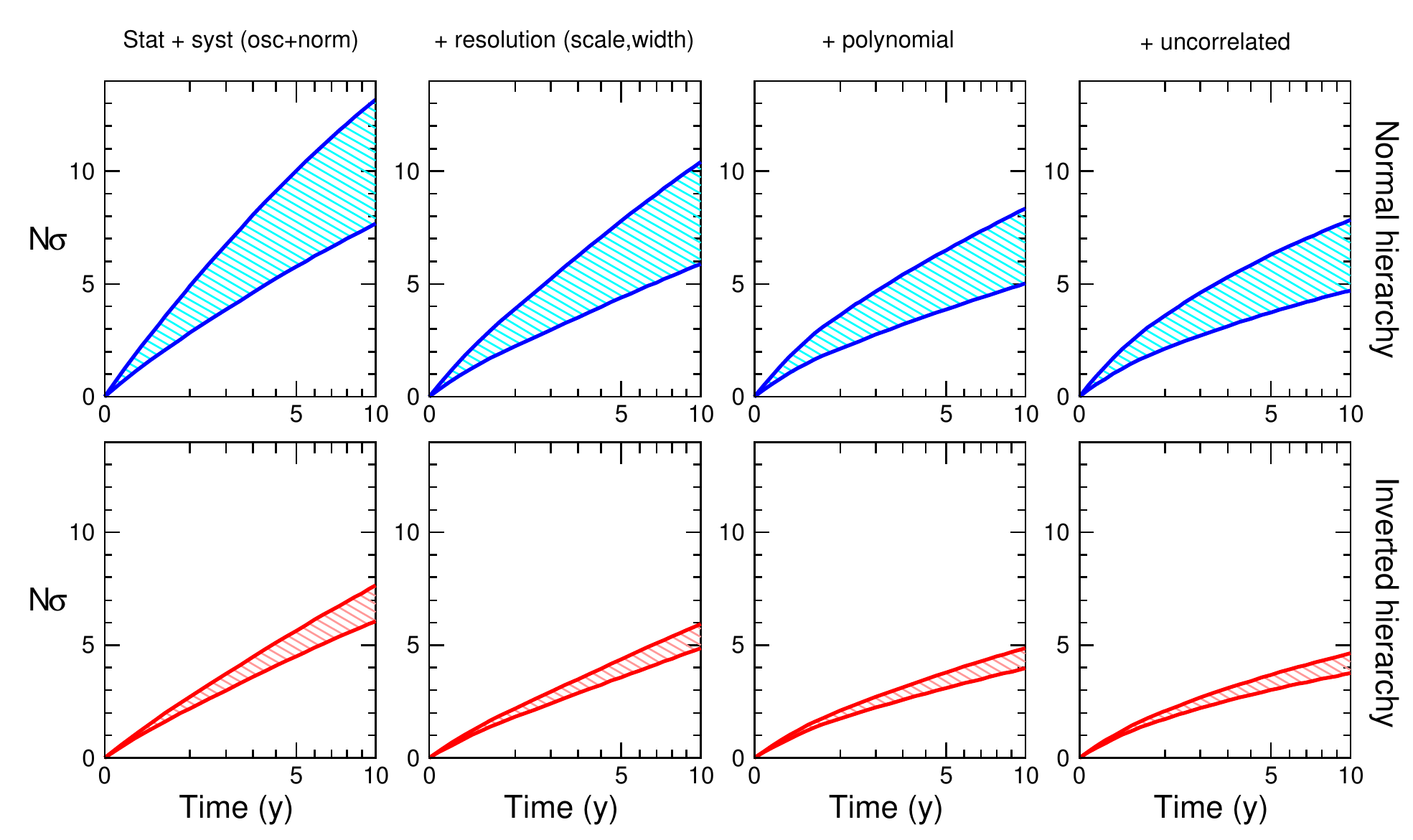}
\caption{As in Fig.~5, but for $\sin^2\theta_{23}|_\text{true}\in [0.46,\,0.54]$. 
\label{Fig_07}}
\end{figure}

Figure~8 shows, in each panel, the fitted value $\sin^2\theta^\text{fit}_{23}$ (at 1, 2 and $3\sigma$) as a function of the true value
$\sin^2\theta^\text{true}_{23} \in [0.4,\,0.6]$, for the four possible cases where the true and tested hierarchies coincide or not. 
The results are obtained in a representative scenario with 5 years of PINGU data, and with polynomial and uncorrelated shape errors at the 1.5\% level. In order to understand qualitatively such results, we stress that 
most of the hierarchy information (via matter effects)
and octant-asymmetric information is embedded in the
$\mu\leftrightarrow e$ flavor oscillation channel, whose amplitude grows with $\sin^2\theta_{23}$. This 
information is enhanced in normal hierarchy, where matter effects are stronger for neutrinos, characterized 
by a larger cross section than antineutrinos.

\begin{figure}[t]
\centering
\includegraphics[height=10cm]{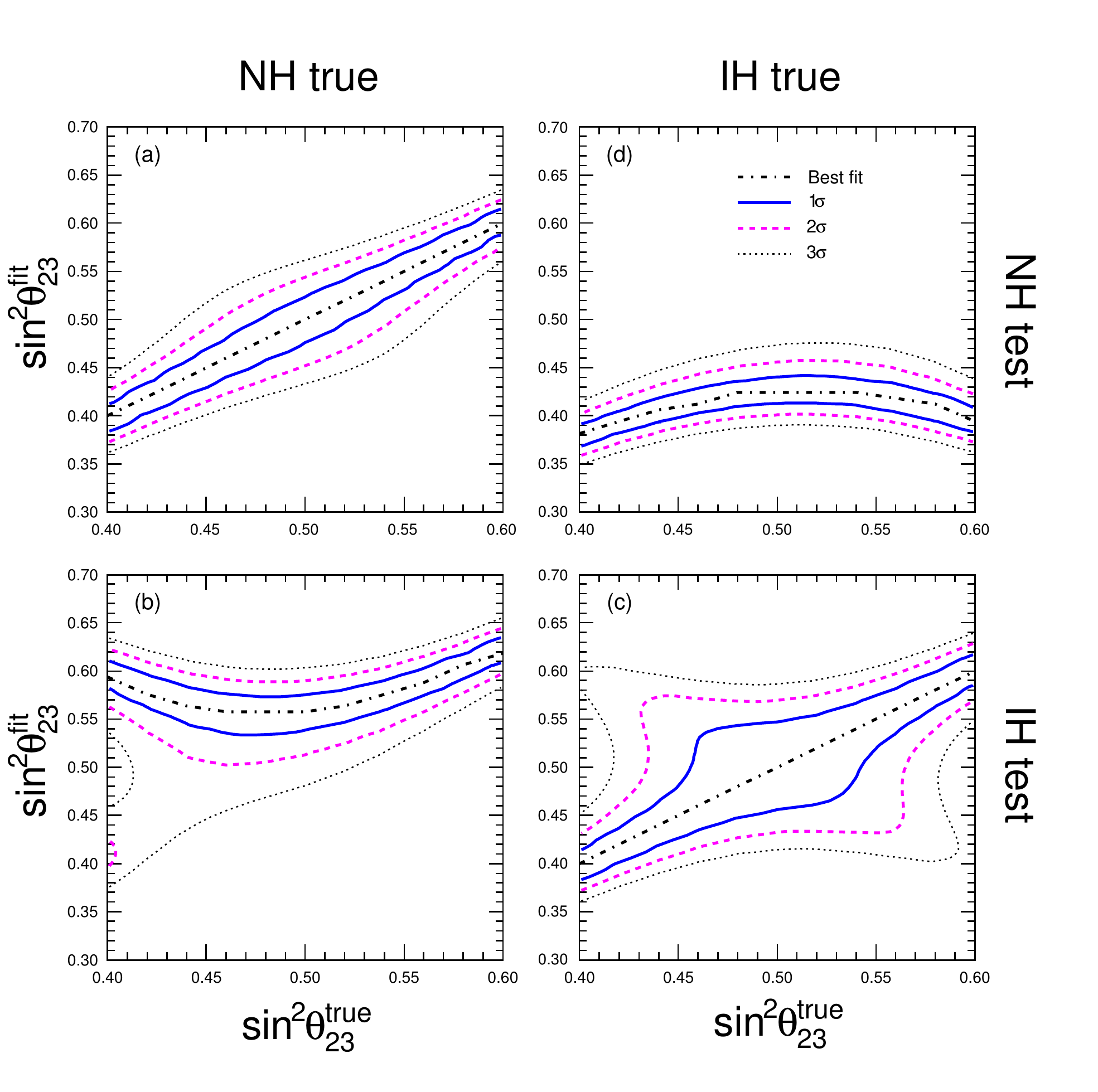}
\caption{Fitted value $\sin^2\theta^\text{fit}_{23}$ (at 1, 2 and $3\sigma$) versus the true value
$\sin^2\theta^\text{true}_{23}$, for the four possible cases where the test hierarchy (i.e., the one assumed in the
fit) is either the true or the wrong one:
(a) NH~=~true, NH~=~test; (b) NH~=~true, IH~=~test; (c) IH~=~true, IH~=~test; (d) IH~=~true, NH~=~test. 
\label{Fig_08}}
\end{figure}

In Fig.~8, the panel (a) refers to the case with true normal hierarchy, assumed to be unambiguously determined by PINGU. 
In this case, by construction, the fitted value $\sin^2\theta^\text{fit}_{23}$ coincides with of the true value
$\sin^2\theta^\text{true}_{23}$; the  1, 2 and $3\sigma$ bands provide then the accuracy of the $\sin^2\theta_{23}$
measurement with PINGU data only. The panel (c) shows analogous case for inverted hierarchy, which clearly shows a worsening of
the accuracy for $\sin^2\theta^\text{fit}_{23}$, and a much more pronounced effect of the octant degeneracy.
The panel (b) refers to the case where the true hierarchy is normal, but PINGU is assumed to mis-identify it 
as inverted. In this case, the fitted value $\sin^2\theta^\text{fit}_{23}$ is systematically higher than the true one; 
the reason is that the fitted IH spectrum tries to reproduce the intrinsically larger effects present in the true NH one, by
increasing $\sin^2\theta^\text{fit}_{23}$ as much as as possible. For analogous reasons, the opposite situation occurs 
in the panel (d), where the true hierarchy is inverted, but PINGU is assumed to mis-identify it 
as normal: the fitted value $\sin^2\theta^\text{fit}_{23}$ is then systematically lower than the true one.
Therefore, until the hierarchy is unambiguously determined, the 
determination of $\sin^2\theta_{23}$ may be subject to strong biases in PINGU.

In conclusion, Fig.~7 illustrates 
the importance of prior information on $\sin^2\theta_{23}$ in determining the PINGU sensitivity to the hierarchy,
while Fig.~8 illustrates, vice versa, the importance of prior information on the true hierarchy in 
determining the PINGU sensitivity to $\sin^2\theta_{23}$.%
\footnote{ In this context, the role of the unknown phase $\delta$ is marginal: we have verified that the reconstructed values of $\delta$ are never constrained above the $1\sigma$ level, for any of the PINGU error configurations examined in our fits (not shown).}

\section{Conclusions and perspectives}

In this work we have examined, in the context of the proposed large-volume detector PINGU,
several issues arising in the calculation of energy-angle distributions of atmospheric muon 
and electron events, and in the associated error estimates. In particular, it has been
shown that the imprint of the neutrino mass hierarchy (either normal or inverted) on these
spectra is sensitive to spectral shape variations at the level of (few) percent. This level of
accuracy, not usually needed in fits to the dominant oscillation parameters $(|\Delta m^2|,\theta_{23})$, 
poses unprecedented challenges to atmospheric neutrino experiments probing subdominant
effects with very high statistics.  

In principle, one should try to characterize, at the percent level, all the known 
independent sources of uncertainties  
coming from models of differential atmospheric fluxes and cross sections, as well as from 
detector models used for event reconstruction. 
Breaking down such uncertainties into many separate nuisance parameters is crucial
for a reliable estimate of the spectral ``flexibility,'' which unavoidably affects the hierarchy sensitivity in 
prospective or real data fits.
Furthermore, in the limit of very high statistics, one should also account (via educated guesses) for residual, poorly
known correlated and uncorrelated systematic uncertainties, which may escape a well-defined parametrization. 
Although each of these error sources may contribute to a tiny reduction of the hierarchy sensitivity, their cumulative
effect may become noticeable.  

In this work, we have first analyzed the PINGU sensitivity to the hierarchy in the presence of
the most obvious systematic errors, due to oscillation parameter and normalization uncertainties.
Then we have added, in sequence, plausible shape systematics related to the resolution functions in energy and 
angle, generic polynomial shape deviations at the (few) percent level, and possible uncorrelated 
systematic errors at a comparable level. We have shown that their cumulative effect can induce a 
 non negligible reduction of the PINGU sensitivity to the hierarchy---a result which deserves further
studies, in order to reach more refined and realistic error estimates.  Finally, we have also 
discussed the interplay between the PINGU sensitivities to the hierarchy and to $\theta_{23}$.

In the context of atmospheric neutrino physics, 
a research program aiming at a better evaluation of the shape uncertainties 
of energy-angle spectra would be beneficial not only for PINGU, but for any
future high-statistics atmospheric experiment, where such spectra 
may either embed subleading oscillation signals or provide a background for other
kinds of emerging signals.  In particular, all the issues discussed herein
are expected to become more severe at lower (sub-GeV) neutrino energy thresholds, such as those 
proposed to probe CP violation effects \cite{Supe}.

Refined characterizations and evaluations of (known and unknown, correlated and uncorrelated) 
spectral uncertainties have been already considered  
in other physics contexts, including fits to parton distribution functions \cite{Pump,Ball}  
and precision cosmology data \cite{Kitc,Tayl}. They are now starting to be required also in the
analysis of short-baseline reactor neutrino data \cite{Dwye}. We understand that 
the PINGU collaboration has recently initiated dedicated investigations towards similar objectives, with encouraging  results
\cite{Tyce,ERes}. We think that, in the era of high-statistics precision experiments, 
these efforts should be largely promoted  in the
neutrino physics community, as it was recognized long ago \cite{Kaji}.


\acknowledgments

This work is supported by the Italian Ministero dell'Istruzione, Universit\`a e Ricerca (MIUR)
and Istituto Nazionale di Fisica Nucleare (INFN) through the 
``Theoretical Astroparticle Physics''  projects.
We are grateful to E.\ Resconi and collaborators for useful discussions 
and clarifications about IceCube and PINGU. We thank 
P.\ Corcella for collaboration in the early stages of this work.
Preliminary results have been shown by A.M.\ at {\em Discrete 2014}, 
Fourth Symposium on Prospects in the Physics of Discrete Symmetries (London, UK, 2014),
and by E.L.\ at the International Conference on Massive Neutrinos (Nanyang Technological Univ., Singapore, 2015).


\end{document}